\documentclass[a4paper,journal]{IEEEtran}
\IEEEoverridecommandlockouts
\usepackage{cite}
\usepackage{amsmath,amssymb,amsfonts}
\usepackage{algorithmic}
\usepackage{graphicx}
\usepackage{textcomp}
\usepackage{xcolor}
\def\BibTeX{{\rm B\kern-.05em{\sc i\kern-.025em b}\kern-.08em
    T\kern-.1667em\lower.7ex\hbox{E}\kern-.125emX}}
    
\usepackage{orcidlink}

\def\mymedskip{\vskip\medskipamount}
\def\mymedbreak{\par \ifdim\lastskip<\medskipamount
  \removelastskip \penalty-100 \mymedskip \fi}
\def\myaftermedspace{\par \ifdim\lastskip<\medskipamount
  \removelastskip \penalty55\mymedskip\fi}
\newcommand{\eop}{{\unskip\nobreak\hfil\penalty50
          \hskip2em\hbox{}\nobreak\hfil$\Box$
          \parfillskip=0pt \finalhyphendemerits=0 \par}}
\newenvironment{proof}%
{\mymedbreak{\noindent\bf Proof.\enspace}}{\eop\myaftermedspace}
\newenvironment{proofn}[1]%
{\mymedbreak{\noindent\bf Proof #1.\enspace}}{\eop\myaftermedspace}
{\mymedbreak{\noindent\bf Proof of Theorem~\ref{#1}:\enspace}}{\eop\myaftermedspace}
{\mymedbreak\noindent{\bf Remark:}%
\enspace\rm}%
{\myaftermedspace}
\newtheorem{teor}{Theorem}[section]
\newtheorem{defi}[teor]{Definition}
\newtheorem{fact}[teor]{Fact}
\newtheorem{problem}{Problem}
\newtheorem{exercise}{Exercise}
\newtheorem{examp}[teor]{Example}
\newtheorem{lem}[teor]{Lemma}
\newtheorem{cor}[teor]{Corollary}
\newtheorem{con}[teor]{Conjecture}
\newtheorem{prop}[teor]{Proposition}
\newtheorem{rem}[teor]{Remark}
\newtheorem{cons}[teor]{Construction}
\newcommand{\beq}{\begin{equation}}
\newcommand{\eeq}{\end{equation}}
\newcommand{\beql}[1]{\begin{equation} \label{#1}}
\newcommand{\eeql}{\end{equation}}
\newcommand{\beqa}{\begin{eqnarray*}}
\newcommand{\eeqa}{\end{eqnarray*}}
\newcommand{\beqal}[1]{\begin{eqnarray} \label{#1}}
\newcommand{\eeqal}{\end{eqnarray}}
\newcommand{\beqan}{\begin{eqnarray}}
\newcommand{\eeqan}{\end{eqnarray}}
\newcommand{\bpf}{\begin{proof}}
\newcommand{\epf}{\end{proof}}
\newcommand{\bpfn}[1]{\begin{proofn}{#1}}
\newcommand{\epfn}{\end{proofn}}
\newcommand{\ben}{\begin{enumerate}}
\newcommand{\een}{\end{enumerate}}
\newcommand{\bit}{\begin{itemize}}
\newcommand{\eit}{\end{itemize}}

\newcommand{\bab}{\begin{abstract}}
\newcommand{\eab}{\end{abstract}}
\newcommand{\bke}{\begin{keywords}}
\newcommand{\eke}{\end{keywords}}

\newcommand{\btm}[1]{\begin{teor} \label{#1}}
\newcommand{\etm}{\end{teor}}
\newcommand{\btmn}[2]{\begin{teor}[#1] \label{#2}}
\newcommand{\etmn}{\end{teor}}
\newcommand{\ble}[1]{\begin{lem} \label{#1}}
\newcommand{\ele}{\end{lem}}
\newcommand{\bLe}[1]{\begin{Lemma} \label{#1}}
\newcommand{\eLe}{\end{Lemma}}
\newcommand{\blen}[2]{\begin{lem}[#1] \label{#2}}
\newcommand{\elen}{\end{lem}}
\newcommand{\bpn}[1]{\begin{prop} \label{#1}}
\newcommand{\epn}{\end{prop}}
\newcommand{\bex}[1]{\begin{examp} \label{#1}}
\newcommand{\eex}{\eop\end{examp}}
\newcommand{\bde}[1]{\begin{defi} \label{#1}}
\newcommand{\ede}{\end{defi}}
\newcommand{\bco}[1]{\begin{cor} \label{#1}}
\newcommand{\eco}{\end{cor}}
\newcommand{\bcorn}[2]{\begin{cor}[#1] \label{#1}}
\newcommand{\ecorn}{\end{cor}}
\newcommand{\bcon}[1]{\begin{con} \label{#1}}
\newcommand{\econ}{\end{con}}
\newcommand{\bfa}[1]{\begin{fact} \label{#1}}
\newcommand{\efa}{\end{fact}}
\newcommand{\bpr}[1]{\begin{problem} \label{#1}}
\newcommand{\epr}{\end{problem}}
\newcommand{\bprnn}[1]{\begin{problemnn} \label{#1}}
\newcommand{\eprnn}{\end{problemnn}}
\newcommand{\bprn}[2]{\begin{problem}[#1] \label{#2}}
\newcommand{\eprn}{\end{problem}}
\newcommand{\bexer}[1]{\begin{exercise} \label{#1}}
\newcommand{\eexer}{\end{exercise}}
\newcommand{\bre}[1]{\begin{rem} \label{#1}}
\newcommand{\ere}{\end{rem}}
\newcommand{\bcons}[1]{\begin{cons} \label{#1}}
\newcommand{\econs}{\end{cons}}
\newcommand{\Tm}[1]{Theorem~\protect\ref{#1}}
\newcommand{\Le}[1]{Lemma~\protect\ref{#1}}

\newcommand{\Ex}[1]{Example~\protect\ref{#1}}

\newcommand{\De}[1]{Definition~\protect\ref{#1}}
\newcommand{\Co}[1]{Corollary~\protect\ref{#1}}

\newcommand{\gl}{\lambda}
\newcommand{\gth}{\theta}

\newcommand{\bbF}{{\mathbb{F}}}

\newcommand{\bbZ}{{\mathbb{Z}}}

\newcommand{\bfaa}{\boldsymbol{a}}
\newcommand{\bfb}{\boldsymbol{b}}
\newcommand{\bfc}{\boldsymbol{c}}

\newcommand{\bfe}{\boldsymbol{e}}
\newcommand{\bfg}{\boldsymbol{g}}

\newcommand{\bfu}{\boldsymbol{u}}
\newcommand{\bfv}{\boldsymbol{v}}

\newcommand{\bfx}{\boldsymbol{x}}

\newcommand{\bfB}{\boldsymbol{B}}

\newcommand{\bfG}{\boldsymbol{G}}

\newcommand{\bfM}{\boldsymbol{M}}

\newcommand{\bfI}{\boldsymbol{I}}

\newcommand{\cA}{{\cal A}}
\newcommand{\cB}{{\cal B}}
\newcommand{\cC}{{\cal C}}
\newcommand{\cD}{{\cal D}}
\newcommand{\cE}{{\cal E}}

\newcommand{\cH}{{\cal H}}

\newcommand{\cL}{{\cal L}}

\newcommand{\cP}{{\cal P}}

\newcommand{\cR}{{\cal R}}
\newcommand{\cS}{{\cal S}}
\newcommand{\cT}{{\cal T}}
\newcommand{\cU}{{\cal U}}

\newcommand{\la}{\langle}
\newcommand{\ra}{\rangle}

\newcommand{\AG}{{\rm AG}}
\newcommand{\PG}{{\rm PG}}

\newcommand{\RH}{\cH}
\newcommand{\RHD}{\cD}
\newcommand{\RHs}{\boldsymbol{H}}
\newcommand{\RHDs}{\boldsymbol{D}}

\newcommand{\choice}[5]{
\left\{ \begin{array}{ll} #1, & \mbox{#2};\\
                                   #3, & \mbox{#4}#5
\end{array}
\right. 
}

\begin{document}

\title{Recovery Models for Linear Batch Codes}
\author{%
Baran D\"uzg\"un, 
Henk D.L.~Hollmann\orcidlink{0000-0003-4005-2369}, 
Ago-Erik Riet\orcidlink{0000-0002-8310-6809}, 
Vitaly Skachek\orcidlink{0000-0002-0626-2437}, 
Vladislav Taranchuk \orcidlink{0000-0002-1375-6670}
\thanks{B.~D\"uzg\"un is with Hacettepe University,
Ankara, T\"urkiye, 
baranduzgun@hacettepe.edu.tr}
\thanks{H.D.L.~Hollmann and V.~Skachek are with University of Tartu, Institute of Computer Science,  Tartu, Estonia, \{henk.hollmann,vitaly.skachek\}@ut.ee}
\thanks{A.-E.~Riet is with University of Tartu, Institute of Mathematics and Statistics, Tartu, Estonia, ago-erik.riet@ut.ee}
\thanks{V.~Taranchuk is with Ghent University, Department of Mathematics, Ghent, Belgium, vlad.taranchuk@ugent.be}
}

\maketitle

\begin{abstract}
Various types of recovery models for batch codes have been considered previously, including asynchronous recovery and recovery as arising from Almost Affinely Disjoint (AAD) families. We initiate a systematic investigation of linear batch codes equipped with recovery algorithms. We introduce online, strongly online,  and strongly asynchronous batch codes, investigate their relations with asynchronous batch codes, and introduce $(m,L)$-strongly asynchronous batch codes, which provide particularly simple recovery algorithms. We then study simple batch codes associated with arbitrary bipartite graphs, obtaining graph-theoretic conditions for such codes, and thereby generalizing several known results.
\end{abstract}


\section{Introduction}
Batch codes are methods for storing data in an encoded form that allow the recovery of a batch of original data symbols from coded symbols in disjoint sets of positions.
Batch codes were introduced in~\cite{IKO+04} for load balancing in distributed storage;  PIR codes for Private Information Retrieval were introduced in~\cite{FVY15}. Switch codes, another related notion, were later introduced in \cite{WKC15}, \cite{WSC+13} to improve data throughput in network switches.
We focus on linear batch codes \cite{LS15}, \cite{TS17}, \cite{VY16}, \cite{ZS16}, where data over a finite field is encoded by a linear transformation.

In this paper, we study linear batch codes together with algorithms for choosing their recovery sets. The batch-code property guarantees that any batch of requests can be served by pairwise disjoint recovery sets, but it does not impose any requirements on how these recovery sets are selected. Here we investigate scenarios in which recovery sets are chosen dynamically. Requests may arrive one at a time, so that a recovery set for the current request must be selected without knowledge of future requests. More generally, recovery processes may finish at different times, and a new request may have to be served while some of the previously selected recovery sets are still in use. This motivates the study of stronger variants of batch codes that allow requests to be served under such dynamic constraints.

We introduce online batch codes, which allow requests to be served one by one, without knowledge of future requests. We also study asynchronous batch codes, introduced in~\cite{RST22}, which allow new requests to be served as active recovery processes finish, using recovery sets that are disjoint from those still in use.
In addition, we introduce stronger variants of these notions, 
where no particular strategy for selecting recovery sets is required: any recovery set for the current request may be chosen, provided that it is disjoint from the recovery sets already in use. 
We show that the strongly online and strongly asynchronous properties are in fact equivalent.
To study when a batch code has these properties, we describe recovery processes by means of recovery histories. We characterize online and asynchronous batch codes by appropriate extension properties of such histories and establish the resulting relations between the different recovery models.

Motivated by constructions of batch codes from Almost Affinely Disjoint (AAD) families, we introduce the notion of an $(m,L)$-strongly asynchronous batch code, which strengthens the strongly asynchronous property by bounding the effect of each recovery-set choice on the choices available for subsequent requests. This reduces the verification of strongly asynchronous recovery to a local condition on recovery-set choices. The notion abstracts the recovery mechanism underlying constructions from $L$-AAD families and isolates the properties of these families that are essential for the resulting batch-code property. This allows us to introduce several variants of $L$-AAD families that retain these essential properties and hence yield batch codes through the same $(m,L)$-strongly asynchronous mechanism.

The notions thus introduced form a natural hierarchy, with increasingly strong requirements on the selection of recovery sets. Obviously, every \(t\)-online batch code is a \(t\)-batch code; in addition, we show that every \(t\)-asynchronous batch code is \(t\)-online, every \(t\)-strongly asynchronous batch code is \(t\)-asynchronous, and an \((m,L)\)-strongly asynchronous batch code is \(t\)-strongly asynchronous for \(t=\lceil m/L\rceil\). We also show that the \(t\)-strongly online and \(t\)-strongly asynchronous properties are equivalent, and that all the resulting notions in this hierarchy are genuinely distinct.

Finally, we study systematic binary batch codes that use only simple recovery sets. The nonsystematic part of the generator matrix of such a code naturally defines a bipartite graph, providing a graph-theoretic description of its simple recovery sets. We obtain graph-theoretic conditions characterizing the $(m,L)$-strongly asynchronous property and use these conditions to generalize several earlier constructions of graph-based batch codes. In particular, we extend results based on bipartite graphs without short cycles to a broader class of graphs excluding certain theta subgraphs.

The remainder of the paper is organized as follows. Section II introduces notation and preliminary results. Section III reviews the batch-code framework, including simple recovery sets, and recalls $L$-AAD and $L$-AAD$^{\prime}$ families before introducing the new notion of $L$-AAD$^{*}$ families and stating their batch-code property. Section IV studies request-serving algorithms for batch codes, including online, asynchronous, strongly online, strongly asynchronous, and $(m,L)$-strongly asynchronous batch codes, and investigates their relations. In Section V we describe methods to determine the various batch-code structures admitted by a given full-row-rank matrix, and we employ these methods to present some notion-separating examples. Section VI considers graph-based batch codes and gives a graph-theoretic characterization of the $(m,L)$-strongly asynchronous property. Finally, Section VII presents some conclusions.

%A preliminary version of this work appeared in~\cite{DHR+26}. 
This paper is an extended version of~\cite{DHR+26}, containing additional details, complete proofs, and some extra results.
\section{\label{LSnot}Notation and preliminaries}
For a positive integer~$n$, we write $[n]$ to denote the set $\{1,2,\ldots, n\}$. We denote by~$S_n$ the group of all permutations on~$[n]$.
We will write $\{\{i_1, \ldots, i_k\}\}$ to denote a multiset with elements $i_1, \ldots, i_k$, where repeated elements are allowed; we sometimes refer to such a multiset as a {\em batch\/}. We write $\bbF_q$ to denote the finite field of size~$q$, and we let $\bfe_1, \ldots, \bfe_n$ denote the $n$ unit vectors in~$\bbF_q^n$. Given vectors $\bfu_1, \ldots, \bfu_m$ in a vector space $V$, we denote by $\la \bfu_1, \ldots,\bfu_m\ra$ the span of these vectors in~$V$.
A {\em prefix\/} of a (possibly empty) sequence $s=s_1,s_2,\ldots, s_t$ is a (possibly empty) sequence $s_1, \ldots, s_j$ with $0\leq j\leq t$. In this paper, a {\em graph\/} is a pair $G=(V,E)$ where $V$ is a finite set referred to as the {\em vertices\/} of~$G$ and $E$ is a subset of (unordered) pairs from~$V$ referred to as the {\em edges\/} of~$G$. A graph $G=(V,E)$ is {\em bipartite\/} if the vertex set $V$ can be partitioned as $V=V_1\cup V_2$ so that every edge intersects both $V_1$ and $V_2$ in a single vertex.
A {\em path\/} of {\em length $k$\/} in a graph (also referred to as a {\em $k$-path\/}) is a sequence of edges $\{P_0,P_1\}, \{P_1,P_2\},\ldots, \{P_{k-1},P_k\}$ where $P_0,P_1,\ldots, P_k$ are distinct vertices. The {\em theta graph} $\gth_{k,t}$ consists of two vertices connected by $t$ internally (vertex) disjoint paths of length~$k$, see, e.g., \cite{Con19}. In particular, $\gth_{k,2}=C_{2k}$, the cycle of length~$2k$.
\section{\label{LSbatch}Batch codes}
Batch codes were originally proposed in~\cite{IKO+04} for load balancing in distributed storage systems.  In this paper, we will only be interested in {\em linear primitive multiset batch codes\/}, or, briefly, {\em linear batch codes\/}. The properties of a linear batch code can be described in terms of its {\em encoder matrix\/}~$\bfG$. Here, an important notion is that of a recovery set.
\bde{LDrec}Let $\bfG=[\bfg_1\bfg_2\cdots \bfg_n]$ be a $k\times n$ matrix over~$\bbF_q$ with columns $\bfg_j$ ($j\in[n]$), and let $i\in[k]$. A set $R\subseteq [n]$ is a {\em recovery set\/} for position (or {\em for request\/})~$i$ with respect to the matrix~$\bfG$ if the $i$th unit vector $\bfe_i$ is contained in the span $\la \bfg_r\mid r\in R\ra$. We say that a recovery set $R$ for~$i$ is {\em minimal\/} if no proper subset of~$R$ is a recovery set for~$i$. 

{\em In this paper, every recovery set is assumed to be minimal.\/}
\ede

Recovery sets can be employed to recover a particular information symbol from its encoding.
The following result, in the special case $\bfb=\bfe_i$, was proved in~\cite{LS15}; the more general formulation below was proved in~\cite{OHR+26}. We include the short proof for completeness.
\bpn{LPrec}
Let $\bfG=[\bfg_1,\ldots,\bfg_n]$ be a $k\times n$ matrix
over~$\bbF_q$, let $R\subseteq[n]$, and let $\bfb\in\bbF_q^k$.
Then the value of~$\bfb\cdot\bfaa$ can be recovered from the coordinates
$(c_r\mid r\in R)$ of $\bfc=\bfaa\cdot\bfG$, for every
$\bfaa\in\bbF_q^k$, if and only if
$\bfb\in\la\bfg_r\mid r\in R\ra$.
\epn
\bpf
Put $W=\la\bfg_r\mid r\in R\ra$. If
$\bfb=\sum_{r\in R}\lambda_r\bfg_r$ with $\gl_r\in\bbF_q$ ($r\in R$), then
\[
\bfb\cdot\bfaa=\sum_{r\in R}\lambda_r c_r,
\]
so $\bfb\cdot\bfaa$ can be recovered from the coordinates indexed
by~$R$.

Conversely, suppose that $\bfb\notin W$. Since
$(W^\perp)^\perp=W$, there exists $\bfaa\in W^\perp$ such that
$\bfb\cdot\bfaa\neq0$. For this $\bfaa$ we have
$c_r=\bfaa\cdot\bfg_r=0$ for every $r\in R$. Thus the encodings
of $\bfaa$ and of the zero vector agree in all coordinates indexed
by~$R$, but $\bfb\cdot\bfaa\neq0=\bfb\cdot{\bf 0}$.
Hence $\bfb\cdot\bfaa$ cannot be recovered from those coordinates
for every~$\bfaa$, a contradiction.
\epf
Formally, a linear batch code can now be defined as follows.
\bde{LDlinbatch}
A {\em linear batch code\/} with parameters $[n,k,t]_q$ is a $k \times n$ matrix $\bfG$ over a finite field $\bbF_q$ with rank~$k$, with the property that for every batch (i.e., for every multiset) $\{\{i_1, \ldots, i_t\}\}\subseteq [k]$, there are pairwise disjoint sets $R_1, \ldots, R_t\subseteq [n]$ such that $R_j$ is a recovery set for position $i_j$ ($j\in[t]$). We sometimes refer to such a code as an $[n,k]_q$ $t$-batch code. 
%We say that the batch code $\bfG$ is {\em systematic\/} if $\bfG$ has the form $[\bfI_k \bfB]$, where~$\bfI_k$ is the $k\times k$ identity matrix.
\ede
Note that in the terminology of~\cite{IKO+04}, a linear batch code is a primitive multiset batch code that is linear. Note also that we identify a linear batch code with its encoding matrix.

In what follows, we will also be interested in linear batch codes of a special type.
\bde{LDsysbc}A {\em systematic\/} linear batch code is a {\em systematic\/} generator matrix,  that is, a $k\times n$ matrix of the form $\bfG=[\bfI \bfM]$, over~$\bbF_q$, where $\bfI$ is the $k\times k$ {\em identity matrix\/} and $\bfM$ a $k\times (n-k)$ matrix. Such a code will be called {\em binary\/} if $q=2$ (so if $\bfM$ is a $(0,1)$ matrix).
\ede
We are also interested in recovery sets of a special type. 
\bde{LDsrec}(See \cite[Definition 2]{PV19}.) Let $\bfG=[\bfI \bfM]$ be a $k\times n$ matrix. A recovery set for position $i$ is called {\em simple\/} if it contains exactly one element from the set of column indices $\{k+1, \ldots, n\}$.
\ede
Note that if $\bfG=[\bfI \bfM]$ is a $k\times n$ matrix over~$\bbF_2$ and if $R=\{i_1, \ldots, i_{m}, k+j\}$ is a recovery set for position $i\in[k]$ with $i_1, \ldots,  i_m\in[k]$ and $1\leq j\leq n-k$, then, by minimality of~$R$, we  have $i\notin \{i_1, \ldots, i_m\}$, and the $(k+j)$th column of~$\bfG$ (equivalently, the $j$th column of~$\bfM$) has a~1 precisely in positions $i,i_1, \ldots, i_m$.
 
Next, we recall the notion of an {\em Almost Affinely Disjoint (AAD) family\/} \cite{PV19} and of an {\em AAD${^\prime}$-family\/} and we introduce that of an AAD$^{*}$-family (a slight generalization of AAD and AAD${^\prime}$ first introduced here).
\bde{LDAAD}(See also \cite[Definition 3]{PV19} and \cite[Definition 1]{LPV+21}.) Let $V=\bbF_q^n$ be an $n$-dimensional vector space over a finite field $\bbF_q$. A collection $\cU=\{U_1, \ldots, U_m\}$ of $k$-dimensional subspaces of~$V$, where $n\geq 2k+1$, is called an {\em $L$-AAD (Almost-Affinely Disjoint) family\/} if 
\ben
\item 
$U_1, \ldots, U_m$ are pairwise skew (that is, $U_i\cap U_j=\{{\bf 0}\}$ for $i\neq j$).
\item
For every $i\in [m]$ and every $\bfv\in V$ with $\bfv\notin U_i$, the coset $\bfv+U_i$ intersects at most $L$ of the subspaces in~$\cU$.
\een
\ede
For the definition of an $L$-AAD$^{*}$ family, we relax the constraint that the subspaces are mutually skew; in addition, we do not require that all subspaces have the same dimension as for $L$-AAD$^{\prime}$ families (see \cite{DHL+26}).
\bde{LDAADs}Let $V=\bbF_q^n$ be an $n$-dimensional vector space over a finite field $\bbF_q$. A multiset $\cU=\{\{U_1, \ldots, U_m\}\}$ of subspaces of~$V$ of dimension at least 1 is called an {\em $L$-AAD$^{*}$ family\/} if 
\ben
\item[1*.] Every member $U\in \cU$ intersects at most $L-1$ other members in~$\cU$ non-trivially (counted with multiplicity).
\item[2.]
For every $i\in [m]$ and every $\bfv\in V$ with $\bfv\notin U_i$, the coset $\bfv+U_i$ intersects at most $L$ of the subspaces in~$\cU$ (counted with multiplicity).
\een
\ede
We recall that an $L$-AAD$^{\prime}$ family is an $L$-AAD$^{*}$ family as in~\De{LDAADs} where all the subspaces $U_i$ have the same dimension \cite{DHL+26}.
Note that an $L$-AAD family certainly is an $L$-AAD$^{\prime}$ family, which in turn is an $L$-AAD$^{*}$ family. The notion of an AAD family is from \cite{PV19}; the definition of an $L$-AAD$^{*}$ family here is new.

Given an $L$-AAD$^{*}$ family $\cU=\{\{U_1, \ldots, U_m\}\}$ in an ambient vector space~$V$ of dimension~$n$ over~$\bbF_q$, we construct a binary batch code as follows. Let $\cC=\cC(\cU)$ denote the multiset of all cosets $\bfv+U_i$ of elements from~$\cU$; note that $N:=|\cC|=\sum_{j=1}^m q^{n-\dim U_j}$. 
%
%Now let $\bfG=[\bfI_V \mid \bfM]$ where $\bfI_V$ is the $|V|\times |V|$ identity matrix with rows and columns indexed by elements of~$V$, and $\bfM$ is a matrix with rows indexed by $V$ and columns indexed by~$\cC$, where the column indexed by $C\in \cC$ is the {\em characteristic vector\/} of~$C$. That is, for $\bfv\in V$ and $C\in \cC$,
%\[M_{\bfv,C}=\choice{1}{if $\bfv\in C$}{0}{otherwise}{.}\]
%
Choose an index set $J$, disjoint from~$V$, with $|J|=N$, and write
\[
  \cC=\{\!\{C_j\mid j\in J\}\!\},
\]
where the indices distinguish different occurrences
of the same coset.

Now let $\bfG=[\bfI_V\mid \bfM]$, where $\bfI_V$ is the
$|V|\times|V|$ identity matrix with rows and
columns indexed by $V$, and $\bfM$ is a matrix
with rows indexed by $V$ and columns indexed
by $J$. For $\bfv\in V$ and $j\in J$, put
\[M_{\bfv,j}=\choice{1}{if $\bfv\in C_j$}{0}{otherwise}{.}\]
In this setting, we index the columns of~$\bfG$ by the elements from $V\cup J$.
%, and a simple recovery set for request~$\bfv \in V$ with respect to~$\bfG$ can be specified by a pair $(\bfv,j)$ in $V\times J$ for which $\bfv\in C_j$, representing the recovery set $C\setminus \{\bfv\}\cup \{C\}$.
For $\bfv\in V$ and
$j\in J$ with $\bfv\in C_j$, the corresponding
simple recovery set for $\bfv$ is
\[
  R(\bfv,j)=\{j\}\cup(C_j\setminus\{\bfv\}).
\]
We now have the following. 
%(We will show this later as a consequence of a more general result, but to motivate what comes later we will discuss a direct proof here.)
%
\btm{LTAAD}With the notation as above, the matrix~$\bfG$ is a $[q^n+N,q^n,t]$ binary batch code with $t=\lceil m/L\rceil$. 
%This batch code has the additional properties that (i) only simple recovery sets are employed, and (ii) requests can be satisfied {\em ``online''\/}, where a new request can be served by an arbitrary simple recovery set disjoint from the recovery sets that have been employed to serve previous requests.
Moreover, \(\bfG\) admits an \((m,L)\)-strongly asynchronous structure (see~\De{LDuol}) in which all recovery sets are simple.
\etm
We will prove the above theorem later as a consequence of a more general result, see~\Co{LCAADs}.
\section{\label{LSreq}Request-serving algorithms for batch codes}
A $[n,k,t]_q$ batch code, by definition, requires that we can serve any batch of at most~$t$ requests by pairwise disjoint recovery sets. However, this requirement does not tell us {\em how\/} to choose the recovery sets. In this section, we will introduce and investigate various  possible request-serving algorithms for batch codes, some investigated before, and some new.  

\subsection{Online batch codes}
An {\em online\/} algorithm bases the current action only on knowledge of the past, but not on the future. 
\bde{LDol}An $[n,k,t]_q$ batch code will be called {\em online\/} if it is possible to serve any request sequence $i_1, \ldots, i_t$ {\em one-by-one\/}, where, for every $j\in [t]$, the choice of the recovery set $R_j$ to serve the current request $i_j$ depends only on the 
recovery sets 
$R_1, \ldots, R_{j-1}$ chosen to serve the preceding requests $i_1, \ldots, i_{j-1}$,
but not on the future requests $i_{j+1}, \ldots, i_t$.
\ede

Note that allowing $R_j$ to depend also on the preceding requests $i_1, \ldots, i_{j-1}$ gives an equivalent definition, 
since these provide no additional information relevant to the choice of~$R_j$ once $R_1, \ldots, R_{j-1}$ are known.

We now describe a useful characterization of online batch codes. We need some preparation.
\bde{LDseqserv}Let $\bfG$ be a full-row-rank $k\times n$ matrix over~$\bbF_q$.  A {\em recovery history sequence\/} for~$\bfG$ is a (possibly empty) sequence $R_1, \ldots, R_s$ ($s\geq0$), where $R_1, \ldots, R_s$ are mutually disjoint recovery sets for~$\bfG$. For a given positive integer~$t$, we refer to such a sequence as {\em $t$-incomplete\/} if $s<t$ and {\em $t$-complete\/} if $s=t$. 
\ede
\bde{LDseqservprop}
Let $\bfG$ be a full-row-rank $k\times n$ matrix over~$\bbF_q$, let $t$ be a positive integer, and let $\RHs$ be a collection of recovery history sequences for~$\bfG$ of length at most~$t$. 
\ben
\item 
$\RHs$ is called {\em prefix-closed\/} if every prefix of a recovery sequence in~$\RHs$ is again contained in~$\RHs$.
\item
$\RHs$ has the {\em extension property\/} (or, is {\em $t$-extendable\/}) if for every incomplete recovery history sequence $R_1, \ldots, R_s$ in~$\RHs$ (so with $s<t$) and every $i\in [k]$, there exists a recovery set $R\subseteq [n]$ for~$i$ such that $R_1, \ldots, R_s, R$ is again contained in~$\RHs$. 
\item $\RHs$ has the {\em exchange property\/} if for every $t$-complete recovery history sequence $R_1, \ldots, R_t$ in~$\RHs$, for every $i\in[k]$ and for every $j\in [t]$, there is a recovery set $R$ for~$i$ such that the sequence $R_1, \ldots, R_{j-1}, R, R_{j+1}, \ldots, R_t$ is again in~$\RHs$.
\een
\ede
\btm{LTob}An $[n,k,t]_q$ batch code $\bfG$ is $t$-online if and only if there exists a nonempty, prefix-closed and $t$-extendable collection $\RHs$ of recovery history sequences for~$\bfG$. 
\etm
\bpf
According to the definition, formally a $t$-online batch code is characterized by {\em choice-functions\/} $f_j$ ($j=1, \ldots, t$). Here $f_1$ associates with every $i\in [k]$ a recovery set $R\subseteq[n]$ for~$i$, and given a request sequence $i_1, \ldots, i_j$ and mutually disjoint recovery sets $R_1$ for $i_1$, \ldots, $R_{j-1}$ for $i_{j-1}$, the choice function $f_j$ associates the sequence $R_1, \ldots, R_{j-1}$ and the last request $i_j$ with a specific recovery set $R_j$ for $i_j$ disjoint from $R_1, \ldots, R_{j-1}$. Now let $\RHs$ be the collection of all sequences $R_1, \ldots, R_s$ ($s\leq t$) of pairwise disjoint recovery sets that can arise after applying the choice functions $f_1, \ldots, f_s$, for a given request sequence $i_1, \ldots, i_s$.
Obviously, $\RHs$ is nonempty and prefix-closed. Moreover, if $R_1, \ldots, R_{s}\in \RHs$ with $s<t$, then, by definition, there are $i_1, \ldots, i_s$ such that $f_j(R_1, \ldots, R_{j-1}, i_j)=R_j$ for $j=1, \ldots, s$;  so if $i\in[k]$, then $R:=f_{s+1}(R_1, \ldots, R_s,i)$ is a recovery set for $i$ disjoint from $R_1, \ldots, R_s$, and by definition, $R_1, \ldots, R_s, R\in\RHs$. 
So $\RHs$ is also $t$-extendable.

Conversely, given a collection $\RHs$ as in the theorem, we can obviously use that collection to create  choice functions $f_1, \ldots, f_t$ for an online batch code (note that these choice functions need not be unique).
\epf
In what follows, an~$[n,k,t]_q$ online batch code will refer to a pair $(\bfG,\RHs)$ as in \Tm{LTob}; we sometimes refer to such a code as {\em $t$-online\/}; we refer to $\RHs$ as a {\em $t$-online structure\/} for~$\bfG$.

The following result can be used to make checking the conditions in~\Tm{LTob} easier.
\btm{LTasynchseq}
Let $\RHs^*$ be a nonempty collection of recovery history sequences of length~$t$, and let $\RHs$ be the prefix-closure of~$\RHs^*$, the collection of all prefixes of sequences in~$\RHs^*$. If $\RHs^*$ has the exchange property, then $\RHs$ is $t$-extendable.
\etm
\bpf
Suppose that $\RHs^*$ has the exchange property. Let  $R_1, \ldots, R_{j-1}$ in~$\RHs$ with $j\leq t$, and let $i\in [k]$. By definition of~$\RHs$, there is a complete recovery history sequence of the form 
$R_1, \ldots, R_{j-1}, R_j, \ldots, R_t$
in~$\RHs^*$. By the exchange property for~$\RHs^*$, there is a recovery set $R\subseteq[n]$ for~$i$ such that 
$R_1, \ldots, R_{j-1}, R,  R_{j+1},\ldots, R_t$
is again contained in~$\RHs^*$. Since $\RHs$ is prefix-closed, the recovery history sequence
$R_1, \ldots, R_{j-1}, R$
is also contained in~$\RHs$. This proves that~$\RHs$ has the extension property.
\epf
\bco{LCobc}Let $\bfG$ be an $[n,k,t]_q$ batch code, and let $\RHs$ be a collection of recovery history sequences for~$\bfG$. Let $\RHs^*$ denote the collection of complete recovery history sequences in~$\RHs$. If $\RHs^*$ is nonempty and has the exchange property, and if $\RHs$ is the prefix-closure of~$\RHs^*$, then $(\bfG,\RHs)$ is $t$-online.
\eco

Not every $[n,k,t]_q$ batch code $\bfG$ admits a $t$-online structure. 
\bex{LEnotonl}
Let 
\[
\bfG:=
\left(\begin{array}{ccccccc}
1&0&1&0&1&0&1\\
0&1&1&0&0&1&1\\
0&0&0&1&1&1&1
\end{array}\right).
\]
Then~$\bfG$ is a $[7,3,4]_2$ batch code (see, e.g., ~\cite{HKR+22}). 
%
%\[\{1\},\{2,3\},\{4,5\},\{6,7\},\{2,4,7\},\{2,5,6\},\{3,4,6\},\{3,5,7\};\]
%
We claim that $\bfG$ does not admit a 4-online batch code. To see this, consider how to serve the request sequence 1,1,1. The recovery sets of size at most two for request 1 are $\{1\}, \{2,3\}, \{4,5\}, \{6,7\}$. 
%So after serving these three requests, at most two columns remain unused. 
Hence, after serving 1,1,1, at most two columns remain unused. 
On the other hand, in order to be able to serve both possible next requests 2 and 3,
%in order to be able to serve either request 2 or request 3 next, 
at least two columns must remain unused. Hence exactly two columns remain unused, and these must be one of the pairs $\{2,3\}$, $\{4,5\}$, or $\{6,7\}$. But if $\{2,3\}$ remains, request 3 cannot be served; if $\{4,5\}$ remains, request 2 cannot be served; and if $\{6,7\}$ remains, neither request can be served. Thus $G$ does not admit a 4-online structure.
\eex
\subsection{Asynchronous batch codes}
 An asynchronous algorithm allows several processes to operate simultaneously and independently. In the case of asynchronous recovery for a $t$-batch code, up to $t$ recovery processes may thus be active at any time. Once a process finishes, a new request may arrive and is assigned a recovery set disjoint from those still in use. We require that any new request can be served whenever fewer than~$t$ recovery processes are active.
Asynchronous batch codes were first introduced in~\cite[Definition III.1]{RST22}. Our treatment here will be slightly different. 
 
We first describe a useful characterization of asynchronous batch codes, in the same style as our earlier characterization of online batch codes. We need some preparation.
\bde{LDserv}Let $\bfG$ be a full-row-rank $k\times n$ matrix over~$\bbF$. A {\em recovery history\/} for~$\bfG$ is a set of the form 
%$\{(i_1,R_1), \ldots, (i_s,R_s)\}$, where $i_1, \ldots, i_s\subseteq [k]$,
$\{R_1, \ldots, R_s\}$ ($s\geq0$),
where $R_1, \ldots, R_s$ are mutually disjoint recovery sets for~$\bfG$.
%, and where $R_j$ is a recovery set for $i_j$, for all $j\in [s]$. 
In the context of a $t$-batch code, we refer to such a recovery history as {\em incomplete\/} if $s<t$ and {\em complete\/} if $s=t$. 
\ede
We think of a recovery history as representing the recovery sets currently in use. 
\bde{LDsetservprop}Let $\RH$ be a collection of recovery histories of size at most~$t$ for a given $[n,k,t]_q$ batch code~$\bfG$. 
\ben
\item 
$\RH$ is called {\em subset-closed\/} if every subset of a recovery history in~$\RH$ is again contained in~$\RH$.
\item
$\RH$ has the {\em extension property\/} (or, is {\em $t$-extendable\/}) if for every recovery history
$\{R_1, \ldots, R_s\}$ in~$\RH$
with $s<t$ and every $i\in [k]$, there exists a recovery set $R\subseteq [n]$ for~$i$, disjoint from $R_1, \ldots, R_s$, such that 
%$\{(i_1,R_1), \ldots, (i_s,R_s), (i,R)\}$ 
%$\{R_1, \ldots, R_s,R\}$
$\{R_1, \ldots, R_s,R\}$
is again contained in~$\RH$. 
\item
We say that $\RH$ has the {\em exchange property\/} if for every complete recovery history
$\{R_1, \ldots, R_t\}$ in~$\RH$,
every $j\in[t]$, and every $i\in [k]$, there exists a recovery set $R$ for~$i$, disjoint from $R_1, \ldots, R_{j-1}, R_{j+1}, \ldots, R_t$, such that 
%\[\{(i_1,R_1), \ldots, (i_{j-1},R_{j-1}), (i,R),  (i_{j+1},R_{j+1}),\ldots, (i_t,R_t)\}\]
$\{R_1, \ldots,R_{j-1},R,R_{j+1}, \ldots, R_t\}$
is again contained in~$\RH$. 
\een
\ede 
\btm{LTab}An $[n,k,t]_q$ batch code $\bfG$ is asynchronous if and only if there exists a nonempty, subset-closed and $t$-extendable collection $\RH$ of recovery histories for~$\bfG$. 
\etm
\bpf
Again, this follows almost directly from the definitions. 
First, suppose that $\bfG$ is asynchronous. Let $\RH$ be the collection of all
recovery histories that can occur as the set of recovery sets currently in use
during an asynchronous recovery process. Clearly, $\RH$ is nonempty, since it
contains the empty recovery history. Moreover, $\RH$ is subset-closed: if a
recovery history is in~$\RH$, then any subset of it can occur by allowing the
recovery processes corresponding to the other recovery sets to finish.

We claim that $\RH$ is also $t$-extendable. Indeed, suppose that
$\{R_1,\ldots,R_s\}\in\RH$ with $s<t$, and let $i\in[k]$. Since $\bfG$ is
asynchronous, a new request for~$i$ can be served while the recovery processes
using $R_1,\ldots,R_s$ are still active. Hence there is a recovery set~$R$ for
$i$, disjoint from $R_1,\ldots,R_s$, such that
\[
\{R_1,\ldots,R_s,R\}\in\RH.
\]
Thus $\RH$ is $t$-extendable.

Conversely, suppose that there exists a nonempty, subset-closed and
$t$-extendable collection $\RH$ of recovery histories for~$\bfG$. Since
$\RH$ is nonempty and subset-closed, the empty recovery history belongs
to~$\RH$. We maintain the invariant that the collection of recovery sets
currently in use is an element of~$\RH$. If a recovery process finishes,
the corresponding recovery set is removed, and the resulting recovery
history is again in~$\RH$ by subset-closedness. If fewer than~$t$ recovery
processes are active and a new request $i\in[k]$ arrives, the extension
property provides a recovery set~$R$ for~$i$ such that, after adding~$R$,
the resulting recovery history is again in~$\RH$. Thus every new request
can be served whenever fewer than~$t$ recovery processes are active, so
$\bfG$ is asynchronous.
\epf
In what follows, an~$[n,k,t]_q$ asynchronous batch code will refer to a pair $(\bfG,\RH)$ as in~\Tm{LTab}; we sometimes refer to such a code as {\em $t$-asynchronous\/}. We refer to~$\cH$ as a {\em $t$-asynchronous structure\/} for~$\bfG$.

The following result makes checking the conditions in~\Tm{LTab} slightly more efficient. We also have a result similar to~\Co{LCobc}.
\btm{LTasynchprop}
Let $\RH^*$ be a nonempty collection of recovery histories of size~$t$, and let $\RH$ be its {\em subset-closure\/}, that is, $\RH$ is the collection of all subsets of sets in~$\RH^*$. Then $\RH^*$ has the exchange property if and only if~$\RH$ is $t$-extendable. 
\etm
\bpf
First, suppose that $\cH^*$ has the exchange property.
Let
%$S=\{(i_1,R_1), \ldots, (i_{j-1},R_{j-1})\}$ 
$\{R_1, \ldots, R_{j-1}\}$ 
be contained in~$\RH$ with $j\leq t$, and let $i\in [k]$. By definition of~$\RH$, there is a complete recovery history of the form 
%$\{(i_1,R_1), \ldots, (i_{j-1},R_{j-1}), (i_j,R_j), \ldots, (i_t,R_t)\}$ 
$\{R_1, \ldots, R_{j-1}, R_j, \ldots, R_t\}$
in~$\RH^*$. By the exchange property for~$\RH^*$, there is a recovery set $R\subseteq[n]$ for~$i$ such that 
%\[\{(i_1,R_1), \ldots, (i_{j-1},R_{j-1}), (i,R),  (i_{j+1},R_{j+1}),\ldots, (i_t,R_t)\}\]
$\{R_1, \ldots, R_{j-1}, R,  R_{j+1},\ldots, R_t\}$
is again contained in~$\RH^*$. Since $\RH$ is subset-closed, the recovery history 
%\[\{(i_1,R_1), \ldots, (i_{j-1},R_{j-1}), (i,R)\}\] 
$\{R_1, \ldots, R_{j-1}, R\}$
is also contained in~$\RH$. This proves that~$\RH$ has the extension property.

Conversely, suppose that $\RH$ has the extension property. Let $\{R_1, \ldots, R_t\}\in \RH^*$, let $j\in [t]$ and let $i\in [k]$. By definition of~$\RH$, we have that $\{R_1, \ldots, R_{j-1}, R_{j+1}, \ldots, R_t\}$ is contained in~$\RH$, and by the extension property for~$\RH$, there is a recovery set~$R$ for~$i$ such that $\{R_1, \ldots, R_{j-1}, R,R_{j+1}, \ldots, R_t\}$ is contained in~$\RH$, hence also in~$\RH^*$. This shows that $\RH^*$ has the exchange property.
\epf
As a consequence, we have the following.
\bco{LCasynch}Let $\bfG$ be an $[n,k,t]_q$ batch code and let~$\RH$ be a collection of recovery histories for~$\bfG$. Then $(\bfG,\RH)$ is $t$-asynchronous if and only if the subset $\RH^*$ of complete recovery histories in~$\RH$ is nonempty and has the exchange property, and $\RH$ is the subset-closure of~$\RH^*$. 
\eco

Next, we investigate the relation between the notions of online and asynchronous batch codes that we defined above. Interestingly, we have the following.
\btm{LTabcisobc}If $(\bfG,\RH)$ is an $[n,k,t]_q$ asynchronous batch code, then $(\bfG,\RHs)$ is an $[n,k,t]_q$ online batch code, where 
%for some suitable~$\RHs$.
\[\RHs:=\{R_{\pi(1)}, \ldots, R_{\pi(s)} \mid \pi\in S_s, \{R_1, \ldots, R_s\}\in \RH\}.\]
\etm
\bpf
%Define  
We claim that $\RHs$ satisfies the conditions of~\Tm{LTob}.
%, that is, $\RHs$ is prefix-closed and $t$-extendable. 
Obviously, since $\RH$ is nonempty and subset-closed, $\RHs$ is nonempty and prefix-closed. Next we show that $\RHs$ is also $t$-extendable. Indeed, suppose that $R_{\pi(1)}, \ldots, R_{\pi(j-1)}$,
originating from a recovery history $\{R_1, \ldots, R_{j-1}\}$ in~$\RH$, is contained in~$\RHs$, and suppose that $i\in [k]$. Then since $\RH$ is 
$t$-extendable, there exists a recovery set $R$ for~$i$ such that 
\[\{R_1, \ldots, R_{j-1},R\}
\in\RH,\]
and hence $R_{\pi(1)}, \ldots, R_{\pi(j-1)}, R$ is also in~$\RHs$. This proves that $\RHs$ is $t$-extendable. 
\epf
So every $t$-asynchronous batch code admits a $t$-online structure. We now show that $t$-asynchronous is a genuinely stronger notion than $t$-online, in the strong sense that not every matrix~$\bfG$ that admits a $t$-online structure admits a $t$-asynchronous structure.
\bex{LEobcnsbc}Let 
\[
\bfG:=
\left(\begin{array}{ccc}
1&0&1\\
0&1&1
\end{array}\right).
\]
Note that request 1 has recovery sets $\{1\}$ and $\{2,3\}$, and request 2 has recovery sets $\{2\}$ and $\{1,3\}$ (here and elsewhere, we number the $n$ columns with $1, \ldots, n$). 
We claim that $\bfG$ admits a $2$-online structure, with online algorithm:
\ben
\item Serve a first request for 1 by $\{1\}$, and a first request for 2 by $\{2\}$.
\item Serve a second request by a recovery set disjoint from the first recovery set.
\een
For example, after first request 1 served by $\{1\}$, we serve a second request 1 by $\{2,3\}$ and a second request 2 by $\{2\}$. So we have
\begin{align}\RHs:=\{(),&(\{1\}), (\{2\}), (\{1\}, \{2\}), (\{2\},\{1\}),\nonumber\\
&(\{1\},\{2,3\}), (\{2\}, \{1,3\})\}.\nonumber
\end{align}
It is easily checked that the algorithm described above can indeed serve every request sequence, and that the set $\RHs$ of recovery history sequences indeed has the required properties of being prefix-closed and 2-extendable.

However, this matrix~$\bfG$ does not admit a 2-asynchronous structure. Indeed, we necessarily have to serve the requests $1,1$ by the recovery sets $\{1\}, \{2,3\}$; now if recovery through $\{1\}$ is finished first, then a new request 2 cannot be served in combination with the recovery set $\{2,3\}$ still in use.
\eex
\subsection{Strongly online and strongly asynchronous batch codes}
We now discuss strong variations of online and asynchronous codes, which turn out to be identical.
\bde{LDsob}
An $[n,k,t]_q$ {\em strongly online\/} batch code is a pair $(\bfG,\cR)$ where $\bfG$ is an $[n,k,t]_q$ batch code and $\cR\subseteq \cP([n])$ is a collection of recovery sets for~$\bfG$  with the property that the collection~$\RHs(\cR)$ consisting of all the recovery history sequences $R_1, R_2, \ldots, R_s$ with $R_1, \ldots, R_s\in \cR$ and $0\leq s\leq t$ is $t$-extendable (for $s=0$, the sequence should be interpreted as $\Lambda$, the empty sequence).
\ede  
So the pair $(\bfG,\cR)$ is $t$-strongly online if we can serve a sequence of requests $i_1, \ldots, i_t$ {\em one-by-one\/}, by choosing an {\em arbitrary\/} recovery set for position $i_1$ from~$\cR$, then choosing an {\em arbitrary\/} recovery set for position $i_2$ from~$\cR$ that is disjoint from~$R_1$, etc.

Similarly, we can also define a stronger form of asynchronous batch codes.
\bde{LDsab}An $[n,k,t]_q$ strongly asynchronous batch code is a pair $(\bfG,\cR)$, where $\bfG$ is an $[n,k,t]_q$ batch code, and $\cR\subseteq \cP([n])$ is a collection of recovery sets for~$\bfG$
%, referred to as {\em admissible recovery sets\/}, 
with the property that the collection $\RH=\RH(\cR)$ consisting of all recovery histories
%$\{(i_1, R_1), \ldots, (i_s,R_s)\}$ 
$\{R_1, \ldots, R_s\}$ with $R_1, \ldots, R_s\in\cR$ and $0\leq s\leq t$ is $t$-extendable (for $s=0$, the set should be interpreted as $\emptyset$, the empty set).
\ede
Interestingly, we have the following.
\btm{LTsosa}Let $\bfG$ be an $[n,k,t]_q$ batch code and let $\cR$ be a collection of recovery sets for~$\bfG$. Then $(\bfG,\cR)$ is $t$-strongly online if and only if $(\bfG,\cR)$ is $t$-strongly asynchronous.
\etm
\bpf
Obviously, $\RHs(\cR)$ is $t$-extendable  if and only if $\RH(\cR)$ is $t$-extendable. Since $\RHs(\cR)$ is nonempty and prefix-closed and $\RH(\cR)$ is nonempty and subset-closed by definition, the result immediately follows from~\Tm{LTob} and~\Tm{LTab}.
\epf
In what follows, we will refer to a pair $(\bfG,\cR)$ satisfying the equivalent conditions of \De{LDsob} and \ref{LDsab} as a {\em $t$-strongly asynchronous\/} batch code, and to~$\cR$ as a {\em $t$-strongly asynchronous structure\/} for~$\bfG$.

In~\Ex{LEanso} in the next section, we show that $t$-strongly asynchronous is a genuinely stronger notion than $t$-asynchronous, in the strong sense that there exist matrices~$\bfG$ that admit a $t$-asynchronous structure, but not a $t$-strongly asynchronous structure.
\subsection{$(m,L)$-strongly asynchronous batch codes}
Inspired by the type of batch codes obtained from Almost Affinely Disjoint ($L$-AAD) families \cite{PV19, LPV+21, OA22,ADO+23} and $L$-AAD${^\prime}$ families \cite{DHL+26}, and by the graph-based batch codes in~\cite{RST22}, we now introduce a type of strongly asynchronous batch code with an even simpler recovery algorithm. We need some preparation.
\bde{LDservice}Let $\bfG$ be a full-row-rank $k\times n$ matrix. A {\em service\/} for~$\bfG$ is a pair $(i,R)$ of a request $i\in[k]$ and a recovery set $R\subseteq [n]$ for~$i$ in~$\bfG$; we sometimes refer to a service $(i,R)$ for~$\bfG$ as a {\em service for~$i$\/}.  
A service for a given batch $\{\{i_1, \ldots, i_s\}\}$ of requests is a set $\{(i_1,R_1), \ldots, (i_s,R_s)\}$ of services where $R_1, \ldots, R_s$ are mutually disjoint.
\ede
\bde{LDuol}Let $m$ and $L$ be positive integers. A full-row-rank $k\times n$ matrix $\bfG$ over $\bbF_q$ will be called {\em $(m,L)$-strongly asynchronous\/} if there exists a multiset $\cS$ of services for~$\bfG$ with the following properties. 
\ben
\item For every $i\in [k]$, there are at least $m$ services $(i,R)$ (counted with multiplicity) for~$i$ in~$\cS$. 
\item For every $i,j\in [k]$ (possibly with $i=j$), every service $(i,R)$ for $i$ in~$\cS$ excludes at most $L$ services for $j$ in~$\cS$ (counted with multiplicity). Here a service $(i,R)$ {\em excludes\/} a service $(i',R')$ if $R$ and~$R'$ are not disjoint. 
\een
\ede
%$
In what follows, an  {\em $(m,L)$-strongly asynchronous\/} batch code will refer to a pair $(\bfG,\cS)$ as in~\De{LDuol}, and we refer to~$\cS$ as an {\em $(m,L)$-strongly asynchronous structure\/} for~$\bfG$. We sometimes refer to such a code as an $[n,k,m,L]_q$ strongly asynchronous batch code. 
We now show that an $(m,L)$-strongly asynchronous batch code indeed is a batch code, and in fact is a strongly asynchronous batch code. 
%The proof of the next theorem should be compared to the proof of~\Tm{LTAAD}.
%
\btm{LTuol}Let $(\bfG,\cS)$ be an $[n,k,m,L]_q$-strongly asynchronous batch code, and let $\cR$ be the collection of all recovery sets $R$ for which $(i,R)\in \cS$ for some $i\in[k]$. Then $(\bfG,\cR)$ is a $t$-strongly asynchronous batch code with $t=\lceil m/L\rceil$.
\etm
\bpf 
Below, we refer to the elements in~$\cS$ as {\em admissible\/} services.
Let $R_1,\ldots,R_{s-1}\in\cR$ be
mutually disjoint recovery sets, and let $i_s\in [k]$. By definition of~$\cR$, there are $i_1, \ldots, i_{s-1}$ such that $(i_1,R_1), \ldots, (i_{s-1},R_{s-1})$ are in~$\cS$.
By (1), request~$i_s$ has at least~$m$ admissible services in~$\cS$, counted with multiplicity, and by (2), each previously chosen service excludes at most $L$ such occurrences. So if every admissible service for position~$i_s$ is excluded by one of $(i_1,R_1), \ldots, (i_{s-1},R_{s-1})$, then $(s-1)L\geq m$. We conclude that if  $(s-1)L<m$, then there exists an admissible service $(i_s,R_s)$ for $i_s$, and hence $R_1, \ldots, R_{s-1}$ can be extended by~$R_s$. Therefore, $(\bfG,\cR)$ is $t$-strongly asynchronous, where $t$ is the integer for which $(t-1)L<m\leq tL$. 
Finally, $(t-1)L<m\leq tL$ gives $t-1<\frac{m}{L}\leq t$; since $t$ is an integer, we conclude that $t=\lceil m/L\rceil$.
\epf
The above ideas were in fact motivated by the following.
\bco{LCAADs}Let $V$ be an $n$-dimensional vector space over a finite field~$\bbF_q$, and suppose that the multiset $\cU=\{\{U_1, \ldots, U_m\}\}$ of subspaces of~$V$ is an $L$-AAD$^{*}$ family. Let $\cC$ and $\bfG$ be defined as in~\Tm{LTAAD}, with rows and columns of~$\bfG$ indexed by $V$ and $V\cup J$, respectively.
Let $\cS$ denote the collection of services $(\bfv, R)$, where $\bfv\in V$ and $R$ is a simple recovery set for~$\bfv$, that is, a set of column indices of the form 
%$C\setminus \{\bfv\}\cup \{C\}$ with $\bfv\in C\in \cC$.
%$\{j\}\cupC_j\setminus\{\bfv\}$
$\{j\}\cup(C_j\setminus\{\bfv\})$
with $\bfv\in C_j$, $j\in J$.
Then the pair $(\bfG,\cS)$ is a simple $(m,L)$-strongly asynchronous batch code. In particular, \Tm{LTAAD} holds.
\eco
\bpf
First, note that a vector $\bfv\in V$ has $m$ simple recovery sets, namely the sets
\[
R(\bfv,j_i)=\{j_i\}\cup(C_{j_i}\setminus\{\bfv\}),
\qquad C_{j_i}=\bfv+U_i,\quad i\in[m],
\]
where $j_i\in J$ denotes the index of the corresponding occurrence of the coset. Thus condition 1) in~\De{LDuol} holds.

Next, let $\bfv,\bfv'\in V$ be fixed, and let
\[
R=\{j\}\cup(C_j\setminus\{\bfv\}),
\qquad C_j=\bfv+U,
\]
be a simple recovery set for $\bfv$, corresponding to an occurrence of $U$ in $\cU$. To establish condition 2) in~\De{LDuol}, we must show that $R$ meets at most $L$ simple recovery sets
\[
R'=\{j'\}\cup(C_{j'}\setminus\{\bfv'\}),
\qquad C_{j'}=\bfv'+U',
\]
for $\bfv'$, counted with multiplicity.

First, suppose that $\bfv'\in\bfv+U$.
If $R$ meets $R'$, then either $j=j'$,
or the two recovery sets intersect in a column
indexed by $V$.
In the first case, $U'$ is the same occurrence
of $U$ in $\cU$. In the second case,
there exists $\bfx\in R\cap R'\cap V$, and hence
\[
\bfx-\bfv'\in U\cap U'.
\]
Moreover, $\bfx\ne\bfv'$, since $\bfv'$
is excluded from $R'$, so this intersection
is nontrivial. Thus $R$ can meet $R'$
only if $U'$ is the same occurrence of $U$,
or if $U'$ intersects $U$ nontrivially.
By condition 1* of an $L$-AAD$^*$ family,
there are at most $L$ such occurrences
of $U'$ in $\cU$, counted with multiplicity.
Finally, suppose that $\bfv'\notin\bfv+U$. If $R$ meets $R'$,
then
\[
 ((\bfv-\bfv')+U)\cap U'\ne\varnothing.
\]
Since $\bfv-\bfv'\notin U$, condition $2$ of the definition of
an $L$-AAD$^*$ family shows that this can occur for at most
$L$ members $U'$ of $\cU$.
%Finally, if $\bfv'\notin \bfv+U$, then $R$ meets $R'$ only if $\bfv-\bfv'+U$ meets $U'$; since $\bfv-\bfv'\notin U$, by condition 2 of an $L$-AAD$^{*}$ family, this again occurs at most~$L$ times. 

We now conclude that the statement in the corollary holds, and hence  
\Tm{LTAAD} holds. 
\epf
We remark that this result slightly strengthens \cite[Lemma 2]{PV19}, see also \cite[page 3]{LPV+21}, for batch codes obtained from $L$-AAD families. Moreover, it shows that in~\cite[Definition 3]{PV19} and~\cite[Definition 2]{LPV+21}, in order to obtain the desired batch code, the requirement that the subspaces all have the same dimension and form a partial spread can be relaxed. 

In~\Ex{LEsnmLs} in the next section, we show that $(m,L)$-strongly asynchronous with $m>(t-1)L$ is a genuinely stronger notion than $t$-strongly asynchronous, in the strong sense that there exist matrices~$\bfG$ that admit a $t$-strongly asynchronous structure, but not an $(m,L)$-strongly asynchronous structure with $m>(t-1)L$.
\section{\label{LSex}Finding Recovery Structures: Pruning Algorithms and Separation Results}
Throughout this section, $\bfG$ is a full-row-rank $k\times n$ matrix over~$\bbF_q$, and $t$ is a positive integer. We write $\cR(\bfG)$ to denote the collection of all (minimal) recovery sets for~$\bfG$. Recall that a $t$-online batch code structure for~$\bfG$ is a collection $\RHs$ of recovery history sequences of length at most~$t$ that is nonempty, prefix-closed, and $t$-extendable. Similarly, a $t$-asynchronous batch code structure for~$\bfG$ is a collection $\RH$ of recovery histories of size at most~$t$ that is nonempty, subset-closed, and $t$-extendable. Both structures can be seen as subgraphs of an acyclic {\em directed graph\/} or  {\em digraph\/}, and code construction can be seen as applying {\em pruning\/} on this digraph. We now explain the idea. 

%Let $\cR$ denote the collection of all recovery sets of~$\bfG$. 
The digraph $\RHDs(\bfG)$ has as vertices the set $\RHs(\bfG)$ of all the recovery history sequences for~$\bfG$ (see~\De{LDseqserv}), with directed edges defined by extension, that is, with an edge $R_1, \ldots, R_s\longrightarrow R_1,\ldots, R_s,R_{s+1}$ whenever $R_1,\ldots, R_s,R_{s+1}$ is a vertex in~$\RHDs(\bfG)$ with $s\geq0$; the digraph $\RHDs(\bfG)$ has the empty sequence $\Lambda$ as its only {\em source\/} (i.e., vertex without predecessor). Similarly, the digraph $\RHD(\bfG)$ has as its vertices the set $\RH(\bfG)$ of all the recovery histories for~$\bfG$ (see~\De{LDserv}), with an edge $\{R_1, \ldots, R_s\}\longrightarrow \{R_1,\ldots, R_s,R_{s+1}\}$ whenever $\{R_1, \ldots, R_s,R_{s+1}\}$ is a vertex in~$\RH(\bfG)$ with $s\geq0$; the digraph $\RHD(\bfG)$ has the empty set~$\emptyset$ as its only source. A vertex of length (or of size) $t$ is said to be {\em at level\/} $t$ in the digraph. 
%Note that the digraph $\RHDs(\bfG)$ is in fact a {\em directed rooted tree\/}.
Note that the digraph $\RHDs(\bfG)$ of recovery history sequences is a directed rooted tree, whereas the digraph $\RHD(\bfG)$ of recovery histories is, in general, only a rooted acyclic digraph.

\subsection*{Pruning}
{\em Pruning\/} refers to an operation on an acyclic digraph where we remove an ``undesirable'' vertex $v$ together with all the vertices reachable from~$v$ (that is, every vertex on a directed path from~$v$). 
%Fix a positive integer $t$. 
For a subgraph $\RHDs$ of~$\RHDs(\bfG)$, we say that a vertex $v=R_1, \ldots, R_s$ in~$\RHDs$ is {\em $i$-extendable\/} if there is a recovery set $R\in \cR(\bfG)$ for~$i$ such that $w=R_1,\ldots, R_s,R$ is again a vertex in~$\RHDs$. Similarly, for a subgraph $\RHD$ of $\RHD(\bfG)$, we say that a vertex $v=\{R_1, \ldots, R_s\}$ in~$\RHD$ is {\em $i$-extendable\/} if there is a recovery set $R\in \cR(\bfG)$ for~$i$ such that $w=\{R_1,\ldots, R_s,R\}$ is again a vertex in~$\RHD$. In both cases, an undesirable vertex in the subgraph is defined as a vertex at level $t-1$ or less that is not $i$-extendable in the subgraph, for some $i\in [k]$. Observe that if a subgraph $\RHDs$ of~$\RHDs(\bfG)$ or a subgraph $\RHD$ of $\RHD(\bfG)$ contains an undesirable vertex, then the vertex set of that subgraph cannot be a $t$-online structure or a $t$-asynchronous structure, since in that case the vertex set would not be $t$-extendable; moreover, since such a structure is prefix-closed (or subset-closed), no vertex in the subgraph of vertices reachable from that vertex can be contained in such a structure either.

So for a subgraph $\RHDs$ of the digraph $\RHDs(\bfG)$, or for a subgraph $\RHD$ of the digraph $\RHD(\bfG)$, with maximum depth~$t$, if the subgraph contains an undesirable vertex $v$ (i.e., a vertex $v$ at level $t-1$ or less that is not $i$-extendable in the subgraph), then we {\em prune\/} that vertex: we remove $v$ and all the vertices reachable from~$v$ from the sub-digraph. 
 
After pruning this vertex, it is possible that there are still other vertices that are not desirable; also, a predecessor of $v$ that was $i$-extendable, but with $v$ as its only $i$-extension, can now be undesirable because of the removal of~$v$. We keep on pruning vertices until no more undesirable vertices are left. 

A subgraph of $\RHDs$ (or of $\RHD$) is \emph{$t$-stable} (or, simply, {\em stable\/}, if the value of~$t$ is clear from the context) if it has maximum level~$t$ and it does not contain any undesirable vertices. 
%Note that since repeated pruning is a monotone decreasing operation, every such repeated pruning operation eventually results in a $t$-stable subgraph or in the empty subgraph. 
Since the digraph is finite and each pruning step
removes at least one vertex, repeated pruning
eventually results in a $t$-stable subgraph
or in the empty subgraph.

A $t$-stable subgraph is \emph{$t$-strongly stable} if it is completely determined by its level-$1$ vertices, in the sense that every recovery history sequence (respectively, recovery history) of level at most \(t\) consisting of mutually disjoint level-\(1\) recovery sets is actually present. As shown in Section~\ref{LSreq} (see also \Tm{LTsosa}), both in the online case and in the asynchronous case, a $t$-strongly
stable structure corresponds precisely to a $t$-strongly asynchronous
structure. 

Throughout the pruning procedure, the vertex sets under consideration are prefix-closed in the sequence case and subset-closed in the history case; pruning preserves these properties.
So to find the maximal $t$-online (or the maximal $t$-asynchronous) structure, start with the subgraph $\RHDs_t$ of~$\RHDs(\bfG)$ (or of the subgraph $\RHD_t$ of~$\RHD(\bfG)$) made up of all vertices at level $t$ or less, and repeatedly prune that graph until we reach a stable subgraph. If that stable subgraph is empty, then no $t$-online (or $t$-asynchronous) structure for~$\bfG$ exists. In that case, we decrease $t$ by one and start all over. And if the stable subgraph is nonempty, then this subgraph represents the maximum possible $t$-online or $t$-asynchronous structure. Indeed, pruning never removes a vertex belonging to a \(t\)-extendable prefix-closed (respectively, subset-closed) subgraph. Given a stable subgraph, to find smaller stable subgraphs, we remove one of the vertices on level $t$, and apply the prune-until-stable operation again.

%[This needs reworking...] The pruning-until-stable operation on a directed tree/digraph is relatively fast since no branching is needed, and takes only $O(N)$ operations, where $N$ is the number of vertices of the digraph. This can be achieved by investigating vertices level by level, starting at level $t$ and going downwards. All maximal proper stable subgraphs of a stable digraph occur among the \(O(N)\) stable subgraphs obtained by deleting one \mbox{level-\(t\)} vertex and pruning; hence these candidates can be generated in \(O(N^2)\) operations.
The prune-until-stable operation requires no branching and can be implemented by processing the vertices and their successor incidences from the highest level downwards. In contrast, determining smaller stable subgraphs requires branching; recursively determining all stable subgraphs may require exponentially many branches.

\subsection*{The LP problem}

%Let $t\geq1$ be a positive integer. 
We now describe a method to investigate whether $\bfG$ admits an $(m,L)$-strongly asynchronous structure $\cS$ with $\lceil m/L\rceil \geq t$. Define the {\em support\/} of~$\cS$ in~$\cR(\bfG)$ as $\cR(\cS):=\{R\in \cR(\bfG)\mid \mbox{$(i,R) \in \cS$ for some $i\in [k]$}\}$.
By~\Tm{LTuol}, the support $\cR(\cS)$ of~$\cS$ is a \(\lceil m/L\rceil\)-strongly asynchronous, and hence a \(t\)-strongly asynchronous structure.

So our approach is to first determine all $t$-strongly asynchronous structures $\cR$ for~$\bfG$, a problem that we discussed before, and  then, for each such $\cR$, to determine if there exists an $(m,L)$-strongly asynchronous structure~$\cS$ with support $\cR$ in~$\cR(\bfG)$ for which $m>(t-1)L$.

So let $\cR$ be a $t$-strongly asynchronous structure for~$\bfG$, and let $\cS$ be a $(m,L)$-strongly asynchronous structure for~$\bfG$ for which $\cR(\cS)=\cR$.  Write $\cR_i$ ($i\in [k]$) to denote the collection of recovery sets for $i$ in~$\cR$; write $I(R)$ to denote the set of all $i\in [k]$ for which $R \in \cR_i$. Let $x_{i,R}$ ($i\in [k]$, $R\in \cR_i$) 
denote the multiplicity of the service $(i,R)$ in~$\cS$. Then the conditions on the support of~$\cS$ and the conditions in~\De{LDuol} can be reformulated as
\begin{enumerate}
\item 
$x_{i,R}\in\bbZ_{\geq0}$ \quad for all $i\in [k]$ and $R\in \cR_i$; 
\item
$\displaystyle\sum_{i\in I(R)}x_{i,R}\geq1$ \quad for all $R\in \cR$.
  \item $\displaystyle\sum_{R\in \cR_i} x_{i,R} \geq m$ \quad for all~$i\in[k]$,
  \item $\displaystyle\sum_{\substack{R\in \cR_i \\ R \cap S \neq \emptyset}}
        x_{i,R} \leq L$ \quad for all $S \in \cR$ and all $i\in[k]$.
\end{enumerate}
Note that the last condition must hold for every \(S\in\cR\), since \(\cR(\cS)=\cR\) implies that \((j,S)\) is actually present in \(\cS\) for some \(j\in[k]\); the condition is independent of the value of \(j\).
%
%A simple counting argument shows that after $s$ services have been used, at most $sL$ services for any color~$i$ are excluded; hence if $m > sL$, some service for~$i$ remains available. This yields a $t'$-strongly asynchronous structure with $t' = \lceil m/L \rceil$.
%
Setting $y_{i,R} := x_{i,R}/L$ and $Q := m/L$, the existence of $\cS$ requires feasibility of the following linear program:
\begin{equation}
  \label{eq:LP}
  \begin{aligned}
    \max \quad & Q \\
    \text{such that} \quad
    & \sum_{R \in \cR_i} y_{i,R} \geq Q
      \quad \mbox{for all $i\in[k]$};\\
    & \sum_{\substack{R \in \cR_i\\ R \cap S \neq \emptyset}} y_{i,R} \leq 1
      \quad \mbox{for all $S \in \cR$ and $i\in [k]$}, \\
    & y_{i,R} \geq 0 \quad \mbox{for all $i\in [k]$ and $R\in\cR_i$}.
  \end{aligned}
\end{equation}
%We have slightly enlarged the feasible region by replacing the requirement $y_{i,R}>0$ by $y_{i,R}\geq0$. By continuity, this does not change the supremum of the objective function. Moreover, since all coefficients in the LP are rational, its optimum $Q^*$ is rational. Hence $m/L\leq Q^*$, so that $t\leq  \lceil Q^*\rceil$.
%
In passing to \eqref{eq:LP}, we have omitted the exact-support
conditions
\[
   \sum_{i\in I(R)}y_{i,R}>0
   \qquad (R\in\cR).
\]
It is easily seen that this does not change the supremum of the objective function. 
%Indeed, a feasible solution of the relaxed LP may first be scaled by a factor $1-\varepsilon$, thereby creating slack in all upper-bound constraints, and then perturbed by sufficiently small positive amounts so that the above support conditions hold, while changing the objective value by an arbitrarily small amount. 
Since all coefficients in the LP are rational, its optimum $Q^*$ is rational. Hence
\[
  m/L \leq Q^*,
\]
and therefore
\[
   \left\lceil m/L\right\rceil\leq\lceil Q^*\rceil.
\]

%\subsection*{Dual certificates}
%
To certify that no \((m,L)\)-strongly asynchronous structure with support \(\cR\) and \(m>(t-1)L\) exists, it suffices by LP duality to exhibit a feasible solution to the dual of \eqref{eq:LP} with objective value at most \(t-1\).
Such a certificate is verifiable by direct substitution, without appeal to the full LP machinery.
This technique is employed in Example~\ref{LEsnmLs} below.
\bre{LRLPq=2} In the case where $q=2$, every minimal recovery set uniquely determines the request that it can serve.  Hence the multiset $\cS$ can be considered as a multiset of {\em recovery sets\/}. As a consequence, if $q=2$ then for every $R\in \cR$, we have $|I(R)|=1$, and in the LP \eqref{eq:LP}, the variables $y_{i,R}$ can be replaced by $y_R$. In the examples below, we use these simplifications without further comment.
\ere
\subsection{Three examples}
The following examples employ the pruning technique for finding asynchronous and strongly asynchronous structures for a given matrix; in the second example we also use a dual certificate for the relevant LP-problem to show non-existence of an $(m,L)$-strongly asynchronous structure with $m/L>2$.

Our first example demonstrates that allowing multisets of services genuinely extends the scope of Definition~\ref{LDuol}.
% as illustrated by the following example.
%
\bex{LEsnmLsnom}
Let
\[
\bfG=
\begin{pmatrix}
1&0&1&0&1&0&1&1\\
0&1&1&0&0&1&1&1\\
0&0&0&1&1&1&1&1
\end{pmatrix}.
\]
Note that $\bfG$ admits a $3$-strongly asynchronous structure: indeed, put
\[
\begin{aligned}
\cR_1&=\bigl\{\{1\},\{2,3\},\{6,7\},\{6,8\}\bigr\},\\
\cR_2&=\bigl\{\{2\},\{4,6\},\{5,7\},\{5,8\}\bigr\},\\
\cR_3&=\bigl\{\{4\},\{1,5\},\{3,7\},\{3,8\}\bigr\},
\end{aligned}
\]
where $\cR_i$ consists of recovery sets for request $i$, and let
\[
\cR=\cR_1\cup\cR_2\cup\cR_3.
\]
%A direct check shows that
%, after choosing any $s<3$ mutually disjoint recovery sets from $\cR$, every request $i\in[3]$ still has a recovery set in $\cR_i$ disjoint from the previously chosen sets. Hence $\cH(\cR)$ is $3$-extendable, and therefore $(\bfG,\cR)$ is $3$-strongly asynchronous.
A direct check shows that $\cR$ is $3$-strongly asynchronous. Using the pruning technique, it is easily verified that, up to the symmetry $\bfe_2\leftrightarrow \bfe_3$, which induces $\bfg_2\leftrightarrow \bfg_4$ and $\bfg_3\leftrightarrow \bfg_5$, this 3-strongly asynchronous structure for~$\bfG$ is unique.

We now show that,
if $\cS$ is required to be a {\em set\/} of services, then there is no
$(m,L)$-strongly asynchronous structure on $\bfG$ with $m\geq 2L+1$ (so with $\lceil m/L\rceil\geq3$).
For request $1$, the minimal recovery sets are
\[
\begin{split}
&\{1\},\{2,3\},\{4,5\},\{6,7\},\{6,8\},\\
&\{2,4,7\},\{2,4,8\},\{2,5,6\},
  \{3,4,6\},\{3,5,7\},\{3,5,8\},
\end{split}
\]
and the corresponding lists for requests $2$ and $3$ are obtained
similarly.

For a fixed request $i$, let $m_L$ denote the largest number of
distinct services for $i$ that can be selected in such a way that
each selected service excludes at most $L$ selected services for $i$.
A direct check gives
\[
m_1=4,\qquad m_2=5,\qquad m_3=6,
\]
while $m_L<2L+1$ for $L\ge4$. Since also $m_3<7$, $L\ge3$ is impossible immediately, and it remains only to consider $L=1,2$.

Suppose first that $L=2$. For each request there is a unique
collection of five minimal recovery sets satisfying the same-request
exclusion condition. These collections are
\[
\begin{aligned}
\cS_1&=\bigl\{
 \{1\},\{2,3\},\{4,5\},\{6,7\},\{6,8\}
\bigr\},\\
\cS_2&=\bigl\{
 \{2\},\{1,3\},\{4,6\},\{5,7\},\{5,8\}
\bigr\},\\
\cS_3&=\bigl\{
 \{4\},\{1,5\},\{2,6\},\{3,7\},\{3,8\}
\bigr\}.
\end{aligned}
\]
However, the service with recovery set $\{2,3\}\in\cS_1$ excludes
the three services for request $3$ with recovery sets
\[
\{2,6\},\qquad \{3,7\},\qquad \{3,8\}.
\]
Thus the exclusion bound $L=2$ is violated.

If $L=1$, the selected recovery sets for each fixed request must be
pairwise disjoint. There are precisely $15$ collections of at least
three minimal recovery sets for a fixed request with this property.
Checking the resulting $15^3$ triples of collections shows that in
every case some selected service excludes at least two selected
services for one of the other requests. Hence $L=1$ is impossible
as well.

We conclude that, when $\cS$ is required to be a set, $\bfG$ admits no
$(m,L)$-strongly asynchronous structure with $m\geq2L+1$.

On the other hand, this conclusion no longer holds if $\cS$ is
allowed to be a multiset. Indeed, take the following multisets of
%services, where only the recovery sets are displayed:
recovery sets:
\[
\begin{aligned}
\cS_1&=\{\!\{
 \{1\},\{1\},\{2,3\},\{6,7\},\{6,8\}
\}\!\},\\
\cS_2&=\{\!\{
 \{2\},\{2\},\{4,6\},\{5,7\},\{5,8\}
\}\!\},\\
\cS_3&=\{\!\{
 \{4\},\{4\},\{1,5\},\{3,7\},\{3,8\}
\}\!\}.
\end{aligned}
\]
Every service in
$\cS=\cS_1\uplus\cS_2\uplus\cS_3$ excludes at most two services
of each fixed request, counted with multiplicity. Since each
$\cS_i$ has cardinality five, $\cS$ is a $(5,2)$-strongly
asynchronous structure for~$\bfG$.
Since $\lceil 5/2\rceil=3$, 
this example indeed demonstrates that allowing multisets of services in Definition~\ref{LDuol} genuinely enlarges its scope.
Note that the support of~$\cS$ in $\cR(\bfG)$ is precisely the $3$-strongly asynchronous structure $\cR$ displayed at the beginning of the example.
\eex
The example below shows that not every $t$-strongly asynchronous batch code admits an $(m,L)$-strongly asynchronous structure with $\lceil m/L\rceil\geq t$. Thus the latter notion is strictly stronger. It turns out that the two notions are essentially the same for \(t=2\); they also seem to be essentially the same for \(t=3\) when \(k=3\). 
\bex{LEsnmLs}
Let $\bfG$ be the $4\times 11$ binary matrix 
%whose columns are
%\[ (7,10,10,11,12,12,13,14,15,15,15). \]
\[\bfG=
\left(\begin{array}{ccccccccccc}
1&0&0&1&0&0&1&0&1&1&1\\
1&1&1&1&0&0&0&1&1&1&1\\
1&0&0&0&1&1&1&1&1&1&1\\
0&1&1&1&1&1&1&1&1&1&1
\end{array}\right).
\]
We claim that $\bfG$ admits a 3-strongly asynchronous structure, but not
a $(m,L)$-strongly asynchronous structure for any positive integers $m,L$ with $2L<m$. 
Let $\cD=\RHD_3(\bfG)$ be the recovery-history digraph whose vertices are the recovery histories of size at most~\(3\). There are $88$ vertices at level~$1$, corresponding
to the minimal recovery sets in $\cR(\bfG)$, and $1276$ vertices
at level~$2$, corresponding to pairs of disjoint minimal recovery
sets. In the computations below we record only the numbers of
vertices at levels~$1$ and~$2$.

The symmetry group of $\bfG$ has order $48$. Using its action, the
pruning procedure described above can be carried out efficiently.
A level-$2$ vertex $\{A,B\}$ survives only if it is $i$-extendable
for every $i\in[4]$; equivalently, for every $i$ there is a recovery
set $C$ for request $i$, disjoint from $A\cup B$, such that
$\{A,B,C\}$ is present at level~3 in the current subdigraph.

The first stages of pruning, grouping successive level-\(1\) and level-\(2\) pruning steps, give
\[
   (88,1276)\longrightarrow(76,884)
                  \longrightarrow(44,460).
\]
The resulting subdigraph is stable but not strongly stable. For
example, the recovery sets $\{2,4\}$ and $\{1,3,7\}$ both survive
at level~$1$, while the level-$2$ vertex
\[
   \bigl\{\{2,4\},\{1,3,7\}\bigr\}
\]
does not survive. Hence a strongly stable subdigraph cannot contain
both recovery sets at level~$1$.
We therefore branch by pruning one of these two level-\(1\) vertices. Repeating this branching-and-pruning procedure for the remaining missing level-\(2\) vertices yields, up to symmetry, a unique nonempty \(3\)-strongly stable subdigraph. Its $28$ level-$1$
vertices are as follows.

For $i=1,2,3,4$, let
%$\R_i$ consist of the following recovery sets:
%\[
%\begin{aligned}
\beqa
%{LE3sn3mL1}
\cR_1&=&
 \{2,4\},\{5,7\},\{8,9\},\{8,10\},\{8,11\},%\nonumber
 \\
\cR_2&=&
 \{6,8\},
 \{1,2,9\},\{7,9\},
 \{1,2,10\},\{7,10\},%\nonumber
 \\
 &&\{1,2,11\},\{7,11\},%\nonumber
 \\
\cR_3&=&
 \{3,8\},
 \{4,9\},\{1,5,9\},
 \{4,10\},\{1,5,10\},\\
 && \{4,11\},\{1,5,11\},\\
\cR_4&=&
 \{1,9\},\{4,6,9\},\{3,7,9\},
\{1,10\},\{4,6,10\},\\
&&\{3,7,10\},\{1,11\},\{4,6,11\},\{3,7,11\},
%\end{aligned}
\eeqa
%\tag{1}
%\]
and let $\cR:=\cR_1\cup\cR_2\cup\cR_3\cup\cR_4$. Thus $\cR$ is the level-$1$ vertex set of a $3$-strongly stable
subdigraph, and hence is a $3$-strongly asynchronous structure
for~$\bfG$.  For each $R\in\cR$, deleting the level-$1$ vertex corresponding
to $R$ and applying the pruning procedure leaves the empty
subdigraph. Hence no proper subset of $\cR$ supports a
$3$-asynchronous structure, and in particular no proper subset
is a $3$-strongly asynchronous structure.
%The branching computation above also shows that, up to symmetry, this is the unique $3$-strongly asynchronous structure for~$\bfG$.
Together with the branching computation above, this shows
that, up to symmetry, $\cR$ is the unique $3$-strongly
asynchronous structure for~$\bfG$.

% Proof that q* = 9/5 for Example 5.2
% Paste into paper as needed.

We next show that $\bfG$ admits no $(m,L)$-strongly asynchronous
structure with $m>2L$. Suppose that such a structure~$\cS$ exists.
By~\Tm{LTuol}, its support $\cR(\cS)$ is \(\lceil m/L\rceil\)-strongly asynchronous, and hence 3-strongly asynchronous.
By the preceding computation, after applying a symmetry of $\bfG$
we may therefore assume that 
\[
   \cR(\cS)=\cR.
\]

Let $Q^*$ denote the optimum of~\eqref{eq:LP}. Define
\[
\begin{aligned}
\cT_2={}&\{
 \{1,2,9\},\{7,9\},
 \{1,2,10\},\{7,10\},
 \{1,2,11\},\{7,11\}\},\\
\cT_3={}&\{\{4,9\},\{4,10\},\{4,11\}\},
\end{aligned}
\]
and put $\cT=\cT_2\cup\cT_3$.

For every $S\in\cT$, the blocking constraint in~\eqref{eq:LP}
corresponding to request~$4$ is
\[
  \sum_{\substack{R\in\cR_4\\R\cap S\ne\emptyset}}y_R\le1.
\]
Summing these nine inequalities with weight $1/5$ gives
\[
 \frac15\sum_{R\in\cR_4}y_R
 \bigl|\{S\in\cT:R\cap S\ne\emptyset\}\bigr|
 \le\frac95.
\]
A direct check shows that every $R\in\cR_4$ intersects exactly five
members of~$\cT$. Hence
\[
   \sum_{R\in\cR_4}y_R\le\frac95.
\]
Together with the coverage constraint
\[
   Q\le\sum_{R\in\cR_4}y_R,
\]
this yields
\[
   Q\le\frac95.
\]
Thus $Q^*\le9/5<2$, and consequently no $(m,L)$-strongly
asynchronous structure with $m>2L$ exists.
\eex
%
%\mymedskip
%
%\begin{center}
%\renewcommand{\arraystretch}{1.15}
%\begin{tabular}{ll}
%\hline
%$R \in \mathcal{R}_4$ & Sets in $\cS$ intersecting $R$ \\
%\hline
%$\{1,9\}$     & $\{1,2,9\},\{7,9\},\{1,2,10\},\{1,2,11\},\{4,9\}$ \\
%$\{4,6,9\}$   & $\{1,2,9\},\{7,9\},\{4,9\},\{4,10\},\{4,11\}$ \\
%$\{3,7,9\}$   & $\{1,2,9\},\{7,9\},\{7,10\},\{7,11\},\{4,9\}$ \\
%$\{1,10\}$    & $\{1,2,9\},\{1,2,10\},\{7,10\},\{1,2,11\},\{4,10\}$ \\
%$\{4,6,10\}$  & $\{1,2,10\},\{7,10\},\{4,9\},\{4,10\},\{4,11\}$ \\
%$\{3,7,10\}$  & $\{7,9\},\{1,2,10\},\{7,10\},\{7,11\},\{4,10\}$ \\
%$\{1,11\}$    & $\{1,2,9\},\{1,2,10\},\{1,2,11\},\{7,11\},\{4,11\}$ \\
%$\{4,6,11\}$  & $\{1,2,11\},\{7,11\},\{4,9\},\{4,10\},\{4,11\}$ \\
%$\{3,7,11\}$  & $\{7,9\},\{7,10\},\{1,2,11\},\{7,11\},\{4,11\}$ \\
%\hline
%\end{tabular}
%\end{center}
%
%\mymedskip
%
%
It is not difficult to prove that every asynchronous $t$-batch code is $t$-strongly asynchronous for $t\leq2$. The following example shows that the two notions may be distinct for $t\geq3$. 
\bex{LEanso}
Consider the binary $4\times 10$ matrix
% 1,2,3,4,5,7,8,9,11,15
\[
\bfG=
\begin{pmatrix}
1&0&1&0&1&1&0&1&1&1\\
0&1&1&0&0&1&0&0&1&1\\
0&0&0&1&1&1&0&0&0&1\\
0&0&0&0&0&0&1&1&1&1
\end{pmatrix}.
\]
We claim that $\bfG$ is a $3$-asynchronous batch code, but not a $3$-strongly asynchronous batch code.

For each $i\in[4]$, there are $17$ minimal recovery sets for request~$i$, so that $|\cR(\bfG)|=68$. 
Let $\cD=\RHD_3(\bfG)$ be the recovery-history digraph whose vertices
are the recovery histories of size at most~$3$. We apply the pruning
procedure described above. Recording only the number of surviving level-2 vertices, the successive numbers are
\[
688,\ 446,\ 268,\ 202,\ 124,\ 97,\ 89,\ 76,\ 76.
\]
Thus the pruning process stabilizes with $76$ vertices at level~$2$.
The corresponding surviving level-$1$ vertices are the following
$15$ recovery sets:
\[
\begin{array}{c|cccc}
1&a=\{1\}&b=\{2,3\}&c=\{4,5\}&d=\{7,8\}\\
2&e=\{2\}&f=\{5,6\}&g=\{8,9\}&\\
3&h=\{4\}&i=\{1,5\}&j=\{3,6\}&k=\{9,10\}\\
4&\ell=\{7\}&m=\{1,8\}&n=\{3,9\}&o=\{6,10\}.
\end{array}
\]
Among these $15$ level-$1$ vertices there are $83$ pairs of disjoint
recovery sets. Of the corresponding level-$2$ vertices, $76$ survive;
the seven missing ones are
\beql{LEmiss}
 \{c,d\},\{c,n\}, \{c,o\}, \{d,j\}, \{d,k\}, \{i,j\}, \{m,n\}.
\eeql
Hence the resulting stable subdigraph is not strongly stable.
We claim that $\bfG$ admits no $3$-strongly asynchronous structure.
Suppose otherwise, and let $\cR$ be the level-$1$ vertex set of a
$3$-strongly stable subdigraph. Since the subdigraph obtained above
is the maximal stable subdigraph, $\cR$ is contained in the above
set of $15$ recovery sets.

Moreover, \(\cR\) must contain at least three pairwise disjoint recovery sets for each request, since the sequence consisting of three copies of the same request must be serviceable.

Since a strongly stable subdigraph contains every history formed by mutually disjoint level-1 recovery sets that it contains, \(d\in\cR\) forces \(j,k\notin\cR\).
Hence at most two recovery sets for
request~$3$ can belong to~$\cR$, a contradiction. Thus $d\notin\cR$.
Similarly, if $c\in\cR$, then $n,o\notin\cR$, and hence at most two
recovery sets for request~$4$ can belong to~$\cR$. Thus
$c\notin\cR$. But then only $a,b$ remain for request~$1$, again a
contradiction.

On the other hand, the subdigraph obtained by the pruning process is
nonempty and stable. Hence its vertex set is a $3$-asynchronous
structure for~$\bfG$. Thus $\bfG$ is a $3$-asynchronous batch code,
but not a $3$-strongly asynchronous batch code.

For completeness, we describe an explicit asynchronous recovery
strategy. Partition the $15$ surviving recovery sets into four
classes:
\ben
\item the {\em singletons\/} $a=\{1\},  e=\{2\}, h=\{4\}, \ell=\{7\}$;
\item the {\em good\/} pairs $f=\{5,6\}, g=\{8,9\}, j=\{3,6\}, k=\{9,10\}, n=\{3,9\},o=\{6,10\}$;
\item the {\em doubtful\/} pairs $i=\{1,5\}, m=\{1,8\}$; 
\item the {\em bad\/} pairs $b=\{2,3\}, c=\{4,5\}, d=\{7,8\}$.
\een
Say that two of the above recovery sets are \emph{compatible} if
they are disjoint and their pair is a surviving level-$2$ vertex;
equivalently, if they are disjoint and do not form one of the pairs listed in~(\ref{LEmiss}).
Now the asynchronous strategy is as follows:
{\em When a new request arrives, first try its singleton. If that cannot be used, try a good pair. If no good pair is available, try a doubtful pair. And only as a last resort, use one of the bad pairs. At every stage discard a candidate if it is incompatible with one of the active recovery sets.\/}
It can be checked that this strategy works, showing in another way that this code is 3-asynchronous.
\eex
We do not know an example with \(k=3\), and conjecture that \(k=4\) is minimal for \(t=3\).
\section{Graph-based batch codes}

In this section, we consider recovery sets of a special type, such as those occurring in graph-based batch codes and batch codes obtained from $L$-AAD families. We restrict attention to {\em binary\/} batch codes, so throughout this section, all matrices are over~$\bbF_2$. We need the following definition; simple recovery sets were defined already in~\De{LDsrec}, and repeated here for convenience.
\bde{LDsrec1}(See \cite[Definition 2]{PV19}.) Let $\bfG=[\bfI_k \bfB]$ be a systematic (binary) $k\times n$ matrix, where $\bfI_k$ is the $k\times k$ identity matrix and $\bfB=(B_{i,j})$ is a binary $k\times (n-k)$ matrix. A recovery set for position $i$ in $\bfG$ is called {\em simple\/} if it contains exactly one element from the set of column indices $\{k+1, \ldots, n\}$. 
We say that a $t$-batch code $\bfG$ is {\em simple\/} if it is systematic and every batch of $t$ requests can be served using only simple recovery sets.
\ede
Simple batch codes $\bfG=[\bfI_k\mid\bfB]$ are best
understood in terms of a bipartite graph associated
with $\bfB$, as introduced in~\cite{RSD+16}.
Choose an index set $J$, disjoint from $[k]$,
with $|J|=n-k$, and regard the columns of
$\bfB$ as indexed by $J$; consequently the columns of $\bfG$ are now indexed by $[k]\cup J$.
Let $\Gamma=\Gamma(\bfB)$ be the bipartite graph
with vertex set $[k]\cup J$ and biadjacency
matrix $\bfB$ (\cite[page 17]{ADH98}). Thus for $i\in[k]$ and $j\in J$, the pair $\{i,j\}$ is an edge of $\Gamma$ if and only if $B_{i,j}\ne 0$.
Write $V=V(\Gamma):=[k]\cup J$.
For $v\in V$, let $\Gamma(v)$ denote the
neighborhood of $v$, and put
$d(v):=|\Gamma(v)|$.

Note that there is a one-to-one correspondence
between edges in $\Gamma$ and simple recovery
sets for $\bfG$, where an edge $\{i,j\}$,
$i\in[k]$, $j\in J$, corresponds to the
simple recovery set
\[
R(i,j):=\{j\}\cup(\Gamma(j)\setminus\{i\})
\]
for $i$. 

We now investigate the batch code properties of the simple batch codes associated with~$\Gamma$. The following lemma characterizes when two simple recovery sets intersect.
\ble{LLrecint}
The edges $\{i,j\}$ and $\{i',j'\}$
($i,i'\in[k]$, $j,j'\in J$) in $\Gamma$
represent intersecting recovery sets if and only if
one of the following holds:
\begin{enumerate}
\item $j'=j$ (this includes the case where
$i'=i$ and the recovery sets are equal).
\item $j'\ne j$, $i'=i$, and there is a
$4$-cycle
$i\sim j\sim i''\sim j'\sim i$
for some $i''\in[k]$ with $i''\ne i$.
\item $j'\ne j$, $i'\ne i$, and there is a
$4$-path
$i\sim j\sim i''\sim j'\sim i'$
in $\Gamma$, for some
$i''\in[k]\setminus\{i,i'\}$.
\end{enumerate}
\ele
\bpf
This follows immediately from the definitions of
$R(i,j)$ and $R(i',j')$.
\epf 
Our main result on (graph-based) simple batch codes is the following.
\btm{LTsub}Let $\bfG=[\bfI_k \bfB]$ be a systematic binary $k\times n$ matrix, let $\Gamma=\Gamma(\bfB)$ be the associated bipartite graph, and let 
$\cS_\Gamma:=\{(i,R(i,j))\mid \mbox{$\{i,j\}$ is an edge in $\Gamma$}\}$.
Then $(\bfG,\cS_\Gamma)$ is a simple $(m,L)$-strongly asynchronous $[n,k,t]_2$ batch code with $t=\lceil m/L\rceil$ 
if and only if the following conditions hold:
\ben
\item For every $i\in [k]$, we have $d(i)\geq m$.
\item For every edge $\{i,j\}$ in~$\Gamma$, there are at most $L-1$ vertices $j'\in J\setminus \{j\}$ contained in a 4-cycle in~$\Gamma$ together with $\{i,j\}$ .
\item 
Given $i,i'\in [k]$ with $i'\neq i$ and an edge $\{i,j\}$ in~$\Gamma$, %such that $i'\notin \Gamma(B)$,  
there are at most $L$ edges $\{i',j'\}$ such that $\{i,j\}$ and $\{i',j'\}$ are the terminal edges of a path in~$\Gamma$ of length 2 or 4.
\een
\etm
\bpf
%Recall that a simple recovery set for $i\in [k]$ takes the form of a  recovery set $R(i,j)$ for a block $B_j\in \cB$ with $i\in \Gamma(B_j)$,  and consists of the ``non-systematic'' column of~$\bfG$ indexed by $J$ together with the ``systematic'' columns of~$\bfG$ corresponding to points in $\Gamma(B_j)\setminus \{i\}$. 
%
Recall that a simple recovery set for $i\in[k]$
has the form
$R(i,j)=\{j\}\cup(\Gamma(j)\setminus\{i\})$
for some $j\in J$ adjacent to~$i$.

According to \De{LDuol}, $(\bfG,\cS_\Gamma)$ is $(m,L)$-strongly asynchronous
%for the collection $\cS_\Gamma$ in the theorem 
precisely when there are at least $m$ services in~$\cS_\Gamma$ for each $i\in[k]$ and if every recovery set $R(i,j)$ for~$i$ intersects at most $L$ recovery sets for every $i'\in [k]$. The first condition obviously holds precisely when for every $i\in [k]$, we have $d(i)\geq m$, which is condition~1 in the theorem. For the second condition, we will use \Le{LLrecint}. So consider a recovery set $R(i,j)$ corresponding to the edge $\{i,j\}$ in~$\Gamma$, for $i\in[k]$ and  $j\in J$. This recovery set intersects itself in $j$, and intersects another recovery set $R(i,j')$ with $j'\neq j$ for~$i$ in a point~$i'\in[k]$ precisely when $i- j - i'-j'-i$ is a 4-cycle containing the edge $\{i,j\}$. So the second condition holds for recovery sets for~$i$ (so with $i'=i$ in condition 2 of \De{LDuol}) precisely when condition~2 in the theorem holds. 

Finally, let $i'\in [k]$ with $i\neq i'$. 
By~\Le{LLrecint}, a recovery set $R(i',j')$ intersects $R(i,j)$ if and only if either $j'=j$, in which case $\{i,j\}$ and $\{i',j'\}$ are the terminal edges of a path of length 2, or $j'\neq j$ and there is a path of length~4 having $\{i,j\}$ and $\{i',j'\}$ as its terminal edges. Hence $R(i,j)$ intersects at most $L$ recovery sets for request~$i'$ if and only if there are at most $L$ edges $\{i',j'\}$ satisfying the condition in part 3. This proves the result.
\epf
\bco{LCsub}Let $\Gamma$ be a bipartite graph with bipartition
classes $\cA$ and $\cB$, where $|\cA|=k$.
Identify $\cA$ with $[k]$, and put $J:=\cB$.
Let $\bfB$ be the biadjacency matrix of
$\Gamma$, with rows indexed by $[k]$
and columns indexed by $J$.
Suppose that $\Gamma$ satisfies the three
conditions from \Tm{LTsub}.
Let $\bfG=[\bfI_k\mid\bfB]$, and let
$\cS_\Gamma$ be as in~\Tm{LTsub}.
Then $(\bfG,\cS_\Gamma)$ is an
$(m,L)$-strongly asynchronous $[n,k,t]_2$
batch code with $n=k+|\cB|$ and
$t=\lceil m/L\rceil$.
\eco
Specializing \Tm{LTsub} to $C_4$-free bipartite graphs gives the following.
\bco{LCnoC4}
(i) Let $\Gamma=(\cP\cup\cB,\cE)$ be a bipartite graph with partition classes $\cP$ and $\cB$. Identify $\cP$ with $[k]$, where $k:=|\cP|$,
and put $J:=\cB$.
Let $\bfG=[\bfI_k\mid\bfB]$, where $\bfB$
is the biadjacency matrix of $\Gamma$,
with rows indexed by $[k]$ and columns
indexed by $J$. Suppose that $\Gamma$ does not contain a~$C_4$. Let $\cS_\Gamma$ be the collection of services
defined in~\Tm{LTsub}. Then $(\bfG,\cS_\Gamma)$ is a simple $(m,L)$-strongly asynchronous batch code if and only if
\ben
\item
$d(P)\geq m$ for all $P\in \cP$.
\item
$\Gamma$ does not contain an edge $\{P,B\}$ with $B\in \cB$ together with a $\gth_{3,L+1}$ disjoint from~$P$ having~$B$ as an endpoint.
\een
(ii) In particular, if $\Gamma$ is a $\{C_4,\gth_{3,L+1}\}$-free bipartite graph and if every vertex $P\in \cP$ has degree at least~$m$, then $(\bfG,\cS_\Gamma)$ is a simple $(m,L)$-strongly asynchronous $[|\cP|+|\cB|,|\cP|,t]_2$ batch code with $t=\lceil m/L\rceil$.

\eco
\bpf
(i) Suppose that $\Gamma$ does not contain a $C_4$. Then condition~2) in~\Tm{LTsub} is automatically satisfied. Moreover, if $P'\in\Gamma(B)$, then the path $P\sim B\sim P'$ is the only path of length 2 or 4 having $\{P,B\}$ and an edge incident with $P'$ as terminal edges, since any such path of length 4 would create a $C_4$.
In addition, if there are two distinct paths of length 3 between a vertex $P\in \cP$ and a vertex $B\in \cB$, then these paths are automatically internally (vertex) disjoint. We conclude that the three conditions in~\Tm{LTsub} are satisfied if and only if $d(P)\geq m$ for all $P\in \cP$ and whenever $\{P,B\}$ is an edge with $P\in \cP$ and $B\in \cB$, there are no $L+1$ internally disjoint paths not containing~$P$ from a vertex $P'\in \cP$ to~$B$.

(ii) This follows immediately from part (i).
\epf
Note that for $L=1$, part (ii) in~\Co{LCnoC4} generalizes~\cite[Theorem 1]{RSD+16}.

Now ~\cite{DRT25} (see also \cite{FT26}) presents a construction of a bipartite $\{C_4, \gth_{3,L+1}\}$-free graph~$\Gamma_{S,n,q}(\cP,\cL)$ from an {\em $(L+1)$-line-evasive\/} set $S$ of~$\PG(n,q)$, a set $S$  of points in~$\PG(n,q)$ with the property that every line in~$\PG(n,q)$ intersects $S$ in at most $L+1$ points; here $\cP$ has size $q^{n+1}$, with any vertex in~$\cP$ of degree $|S|$, and $\cL$ has size $|S|q^{n}$, with any vertex in~$\cL$ of degree~$q$. So by~\Co{LCnoC4}, we obtain a simple $(|S|,L)$-strongly asynchronous $[q^{n}(q+|S|), q^{n+1}]_2$ $t$-batch code with $t= \lceil |S|/L \rceil$.

Recently, several constructions and bounds for bipartite $\gth_{k,\ell}$-free (see, e.g., \cite{Con19}, \cite{JMY22}) and $\{C_4, \gth_{3,L+1}\}$-free graphs \cite{DRT25}, \cite{FT26} have appeared in the literature. We need some terminology. An {\em $(r,s)$-set\/} is a set $K$ in~$\PG(n,q)$ such that every $s$-space in~$\PG(n,q)$ meets $K$ in at most $r$ points. An $(s,r)$-subspace evasive set  \cite{PV19} 
is a set $K$ in~$\AG(n,q)$ such that every $s$-flat in~$\AG(n,q)$ meets $K$ in at most  $r$ points; a $(1,r)$-subspace-evasive set is commonly referred to as an $r$-line-evasive set. Identifying $\AG(n,q)$ with the complement of a hyperplane
in $\PG(n,q)$, every projective $s$-space meets the affine
part either in an affine $s$-flat or in the empty set.
Consequently, every $(s,r)$-subspace-evasive set in
$\AG(n,q)$ is an $(r,s)$-set in $\PG(n,q)$, see~\cite{FT26}. For our purposes, the strongest result is obtained by using a construction from~\cite{FT26}. This yields the following result.
\btm{LTlineev}Let $n\geq 2$ and put
\[
r_n:=\min\left\{r\geq 2:\binom{r+1}{2}\geq n\right\}
=\left\lceil\frac{\sqrt{8n+1}-1}{2}\right\rceil.
\]
For every prime power $q>r_n$, there exists a simple $(q^{n-1}, r_n-1)$-strongly asynchronous $[q^{n+1}+q^{2n-1},q^{n+1},t]_2$ batch code with $t=\lceil q^{n-1}/(r_n-1)\rceil =\Omega(q^{n-1}/\sqrt{n})$.
\etm
\bpf
We employ the construction from~\cite[Proof of Theorem 1.1, pages 7--8]{FT26}. Let $r$ be an integer with
$2\leq r<q$, put
\[
d=d_r:=\binom{r+1}{2},
\]
and let $k$ be the integer such that
\[
d(k-1)<n\leq dk.
\]
The construction in~\cite{FT26} gives a partition of
$\AG(d,q^k)$ into $q^k$ $r$-line-evasive sets. After field
reduction, this yields a partition of $\AG(dk,q)$ into $q^k$
$r$-line-evasive sets. Since $n\leq dk$, we may regard
$\AG(n,q)$ as an affine subspace of $\AG(dk,q)$, hence this partition
induces a partition of~$\AG(n,q)$ into at most $q^k$ nonempty $r$-line-evasive sets. Hence one of these sets has size at least~$q^{n-k}$.

Now take $r=r_n$. Then $d_r\geq n$, so in the above we may take $k=1$. Thus there
exists an $r_n$-line-evasive set $S$ in $\AG(n,q)$ of size at
least $q^{n-1}$. By taking a subset if necessary, we may assume
that $|S|=q^{n-1}$.
Identifying $\AG(n,q)$ with the complement of a hyperplane
in $\PG(n,q)$, we may regard $S$ as an $r_n$-line-evasive
set in $\PG(n,q)$, see~\cite{FT26}.

Finally, as mentioned above (see~\cite{DRT25}),
the linear representation of $S$ produces a
$\{C_4,\theta_{3,L+1}\}$-free bipartite graph
$\Gamma=(\cP\cup\cB,\cE)$ with $L=r_n-1$, where
\[
|\cP|=q^{n+1},\qquad
|\cB|=|S|q^n=q^{2n-1},
\]
and
\[
d(P)\geq m:=|S|=q^{n-1}
\quad\text{and}\quad
d(B)=q
\]
for all $P\in\cP$ and $B\in\cB$. The result now follows from~\Co{LCnoC4}.
\epf

\section{Conclusions}
We have initiated the investigation of batch codes equipped with request-serving algorithms. We introduced $t$-online batch codes, which can serve up to $t$ requests as they arrive, and $t$-asynchronous batch codes, which allow up to $t$ recovery processes to run simultaneously and independently. We also studied $t$-strongly asynchronous batch codes, 
for which recovery sets can be chosen freely from a prescribed collection, subject only to being disjoint,
and the notion of $(m,L)$-strongly asynchronous batch codes. These notions form a strict hierarchy: an $(m,L)$-strongly asynchronous batch code is $t$-strongly asynchronous for $t=\lceil m/L\rceil$, while $t$-strongly asynchronous implies $t$-asynchronous, which implies $t$-online and hence $t$-batch.
Moreover, we showed that all the notions in this hierarchy are genuinely distinct, in the strong sense that there are matrices that admit one of these structures, but do not admit the next stronger structure. 

Motivated by recent constructions from Almost Affinely Disjoint Families, we introduced $L$-AAD* families and showed that an $L$-AAD* family of size $m$ yields a binary $(m,L)$-strongly asynchronous batch code. We also characterized simple $(m,L)$-strongly asynchronous batch codes in terms of forbidden configurations in their associated bipartite graphs. 
These results generalize and improve various previous results on $L$-AAD and $L$-AAD${^\prime}$ families and on graph-based batch codes. By combining a recent construction of $\{C_4,\theta_{3,L+1}\}$-free graphs from line-evasive sets with a recent construction of large line-evasive sets, we obtained simple $[q^{n+1}+q^{2n-1},q^{n+1}]_2$ $t$-batch codes with $t=\lceil q^{n-1}/(r_n-1)\rceil\approx q^{n-1}/\sqrt{2n}$ whenever $n\geq2$ and $q>r_n$, where $r_n=\lceil (\sqrt{8n+1}-1)/2\rceil\approx \sqrt{2n}$.
%;  for sufficiently large $n$, a stronger bound $t=\Omega(q^{n-1}/n)$ is obtained. 
%
\section*{Acknowledgments}
This work is supported in part by the Estonian Research Council grant PRG2531. Baran D\"uzg\"un gratefully acknowledges TÜBİTAK for the support BIDEB-2211-A and BİDEB-2214-A during her visit at the University of Tartu.
%
%\newpage
%

%
\end{document}